\documentstyle[11pt]{article}

\def\AFOUR{%
\setlength{\textheight}{8.5in}%                    
\setlength{\textwidth}{5.75in}%                    
\setlength{\topmargin}{-0.375in}%                  
\hoffset=-.5in%                                    
\renewcommand{\baselinestretch}{1.17}%             
\setlength{\parskip}{6pt plus 2pt}%                
}

\AFOUR                                           %%%   ***

\expandafter\ifx\csname amssym.def\endcsname\relax \else\endinput\fi
\expandafter\edef\csname amssym.def\endcsname{%
       \catcode`\noexpand\@=\the\catcode`\@\space}
\catcode`\@=11
\def\undefine#1{\let#1\undefined}
\def\newsymbol#1#2#3#4#5{\let\next@\relax
 \ifnum#2=\@ne\let\next@\msafam@\else
 \ifnum#2=\tw@\let\next@\msbfam@\fi\fi
 \mathchardef#1="#3\next@#4#5}
\def\mathhexbox@#1#2#3{\relax
 \ifmmode\mathpalette{}{\m@th\mathchar"#1#2#3}%
 \else\leavevmode\hbox{$\m@th\mathchar"#1#2#3$}\fi}
\def\hexnumber@#1{\ifcase#1 0\or 1\or 2\or 3\or 4\or 5\or 6\or 7\or 8\or
 9\or A\or B\or C\or D\or E\or F\fi}

\font\tenmsa=msam10
\font\sevenmsa=msam7
\font\fivemsa=msam5
\newfam\msafam
\textfont\msafam=\tenmsa
\scriptfont\msafam=\sevenmsa
\scriptscriptfont\msafam=\fivemsa
\edef\msafam@{\hexnumber@\msafam}
\mathchardef\dabar@"0\msafam@39
\def\dashrightarrow{\mathrel{\dabar@\dabar@\mathchar"0\msafam@4B}}
\def\dashleftarrow{\mathrel{\mathchar"0\msafam@4C\dabar@\dabar@}}

\def\ulcorner{\delimiter"4\msafam@70\msafam@70 }
\def\urcorner{\delimiter"5\msafam@71\msafam@71 }
\def\llcorner{\delimiter"4\msafam@78\msafam@78 }
\def\lrcorner{\delimiter"5\msafam@79\msafam@79 }
\def\yen{{\mathhexbox@\msafam@55}}
\def\checkmark{{\mathhexbox@\msafam@58}}
\def\circledR{{\mathhexbox@\msafam@72}}
\def\maltese{{\mathhexbox@\msafam@7A}}
\def\circledS{{\mathhexbox@\msafam@73}}

\font\tenmsb=msbm10
\font\sevenmsb=msbm7
\font\fivemsb=msbm5
\newfam\msbfam
\textfont\msbfam=\tenmsb
\scriptfont\msbfam=\sevenmsb
\scriptscriptfont\msbfam=\fivemsb
\edef\msbfam@{\hexnumber@\msbfam}
\def\Bbb#1{{\fam\msbfam\relax#1}}
\def\widehat#1{\setbox\z@\hbox{$\m@th#1$}%
 \ifdim\wd\z@>\tw@ em\mathaccent"0\msbfam@5B{#1}%
 \else\mathaccent"0362{#1}\fi}
\def\widetilde#1{\setbox\z@\hbox{$\m@th#1$}%
 \ifdim\wd\z@>\tw@ em\mathaccent"0\msbfam@5D{#1}%
 \else\mathaccent"0365{#1}\fi}
\font\teneufm=eufm10
\font\seveneufm=eufm7
\font\fiveeufm=eufm5
\newfam\eufmfam
\textfont\eufmfam=\teneufm
\scriptfont\eufmfam=\seveneufm
\scriptscriptfont\eufmfam=\fiveeufm
\def\frak#1{{\fam\eufmfam\relax#1}}

\csname amssym.def\endcsname

\makeatletter
\def\section{\@startsection {section}{1}{\z@}{-3.5ex plus -1ex minus 
 -.2ex}{2.3ex plus .2ex}{\large\sc}}
\def\subsection{\@startsection{subsection}{2}{\z@}{-3.25ex plus -1ex minus 
 -.2ex}{1.5ex plus .2ex}{\normalsize\sc}}
\makeatother

\makeatletter
\@addtoreset{equation}{section}                        

\makeatother

\newcommand{\nc}{\newcommand}
\newcommand{\rnc}{\renewcommand}

\nc{\be}{\begin{equation}}
\nc{\ee}{\end{equation}}
\nc{\bea}{\begin{eqnarray}}
\nc{\eea}{\end{eqnarray}}

\nc{\trac}[2]{{\textstyle\frac{#1}{#2}}}

\nc{\ex}[1]{\mbox{e}^{\,\textstyle#1}}

\nc{\CC}{\Bbb{C}}
\nc{\HH}{\Bbb{H}}
\nc{\PP}{\Bbb{P}}
\nc{\RR}{\Bbb{R}}
\nc{\OO}{\Bbb{O}}
\nc{\ZZ}{\Bbb{Z}}
\nc{\II}{\Bbb{I}}
\nc{\EE}{\Bbb{E}}
\nc{\SS}{\Bbb{S}}

\rnc{\a}{\alpha}
\rnc{\b}{\beta}
\nc{\ab}{\alpha^{*}}
\nc{\al}{\a^{l}}
\rnc{\d}{\delta}
\nc{\ga}{\gamma}
\nc{\la}{\lambda}
\nc{\lal}{\la_{l}}
\nc{\f}{\phi}
\nc{\fb}{\bar{\phi}}
\nc{\p}{\psi}

\nc{\e}{\varepsilon}
\nc{\eb}{\bar{\e}}
\rnc{\c}{\chi}
\nc{\cb}{\bar{\chi}}
\nc{\eps}{\epsilon}
\rnc{\t}{\theta}
\nc{\tb}{\bar{\theta}}
\nc{\om}{\omega}
\nc{\Om}{\Omega}

\rnc{\P}{\Psi}
\nc{\pl}{\P_{L}}
\nc{\pdr}{\P^{\dag}_{R}}
\nc{\Ga}{\Gamma}
\nc{\G}[3]{\Gamma^{#1}_{\;{#2}{#3}}}

\nc{\sig}{\sigma}
\nc{\sk}{\sigma_{k}}
\nc{\sa}{\sigma_{a}}
\nc{\Bb}{\bar{B}}

\nc{\symx}{\circledS}

\nc{\Q}{\bar{Q}}
\nc{\C}{{\cal A}/{\cal G}}                
\nc{\A}{{\cal A}}
\nc{\Ap}{{\cal A}^{+}}
\nc{\RC}{{\cal R}_{\C}}                 
\nc{\RM}{{\cal R}_{\M}}                
\nc{\RX}{{\cal R}_{X}}
\nc{\RY}{{\cal R}_{Y}}

\nc{\ad}{\mathop{\mbox{ad}}\nolimits}
\nc{\tr}{\mathop{\mbox{tr}}\nolimits}
\nc{\Tr}{\mathop{\mbox{Tr}}\nolimits}
\nc{\Det}{\mathop{\mbox{Det}}\nolimits}
\rnc{\det}{\mathop{\mbox{det}}\nolimits}
\nc{\rk}{\mathop{\mbox{rk}}\nolimits}
\nc{\dime}{\mathop{\mbox{dim}}\nolimits}
\nc{\diag}{\mbox{diag}}
\nc{\ra}{\rightarrow}
\nc{\Ra}{\Rightarrow}
\nc{\LRa}{\Leftrightarrow}
\nc{\lra}{\leftrightarrow}
\nc{\ot}{\otimes}
\rnc{\ss}{\subset}
\nc{\ul}{\underline}
\nc{\nul}{\noindent\underline}
\nc{\non}{\nonumber\\}
\rnc{\S}{\Sigma}
\nc{\tp}{2\pi i}
\nc{\del}{\partial}
\nc{\dbar}{\bar{\del}}
\nc{\dx}{\dot{x}}
\nc{\zb}{\bar{z}}
\nc{\g}{\frak{g}}

\nc{\lv}{\log V_{s}}
\nc{\vs}{V_{s}}
\nc{\ls}{\ell_{s}}
\nc{\N}{{\cal N}}
\nc{\M}{{\cal M}}
\rnc{\L}{{\cal L}}
\nc{\F}{{\cal F}}
\nc{\E}{{\cal E}}
\rnc{\P}{{\cal P}}
\nc{\I}{{\cal I}}
\nc{\IIt}{$\widetilde{\mbox{II}}$}
\nc{\gst}{\widetilde{g_{s}}}
\nc{\gsh}{\widehat{g_{s}}}
\nc{\lsh}{\widehat{\ls}}
\nc{\rllh}{\widehat{R_{11}}}
\nc{\lph}{\widehat{\ell_{P}}}
\nc{\mnd}{M_{Nd}}

\nc{\nam}{\nabla_{\mu}}
\nc{\nan}{\nabla_{\nu}}
\nc{\nal}{\nabla_{\la}}
\nc{\ds}{(\d\! *)}

\nc{\va}{v_{A}}
\nc{\vb}{v_{B}}
\nc{\vc}{v_{C}}
\nc{\vam}{v_{A\mu}}
\nc{\van}{v_{A\nu}}
\nc{\val}{v_{A\la}}
\nc{\vbm}{v_{B\mu}}
\nc{\vbn}{v_{B\nu}}
\nc{\vbl}{v_{B\la}}
\nc{\vcm}{v_{C\mu}}
\nc{\vcn}{v_{C\nu}}
\nc{\vcl}{v_{C\la}}

\nc{\mat}[4]{\left(\begin{array}{cc}#1&#2\\#3&#4\end{array}\right)}
 
\nc{\r}[1]{\mathbf{#1}}
\nc{\rb}[1]{\overline{\mathbf{#1}}}

\nc{\gi}{\gamma_{i}}
\nc{\gj}{\gamma_{j}}

\nc{\subs}[1]{{\vspace*{0.5cm}}%
{\noindent\underline{\small\sc #1}}
{\vspace*{0.3cm}}}
\nc{\chap}[1]{{\clearpage}%
\begin{center}%
{\noindent\underline{\large\sc #1}}{\addcontentsline{toc}{section}{#1}}%
\end{center}%
{\vspace*{0.3cm}}}

\newcommand{\ba}{\begin{eqnarray}}
\newcommand{\ea}{\end{eqnarray}}

\begin{document}
%\begin{titlepage}
\begin{flushright}
%IC/01/???\\
{\tt hep-th/0108122}
%\\
%{\sc Version of \today}
\end{flushright}
%\vskip 1in
\begin{center}
{\LARGE\sc Instantons, the Information Metric, \\[.3in]
and the A}{\LARGE d}{\LARGE\sc S/CFT Correspondence}\\[.5in]
{\large\sc Matthias Blau, K.S.\ Narain, George Thompson}\\[.3in]
{\sc The Abdus Salam ICTP\\ Strada Costiera 11\\ 34014 Trieste, Italy}\\
{\small\tt mblau/narain/thompson@ictp.trieste.it}
\end{center}
\vskip .5in
\begin{abstract}
\noindent
We describe some remarkable properties of the so-called Information
Metric on instanton moduli space. This Metric is manifestly gauge and
conformally invariant and coincides with the Euclidean $AdS_{5}$-metric
on the one-instanton $SU(2)$ moduli space for the standard metric on
$\RR^{4}$. We propose that for an arbitrary boundary metric the AdS/CFT
bulk space-time {\em is} the instanton moduli space equipped with the
Information Metric.

\noindent
To test this proposal, we examine the variation of the instanton moduli
and the Information Metric for first-order perturbations of the boundary
metric and obtain three non-trivial and somewhat surprising results: (1)
The perturbed Information Metric is Einstein. (2) The perturbed instanton
density is the corresponding massless boundary-to-bulk scalar propagator.
(3) The regularized boundary-to-bulk geodesic distance is proportional to
the logarithm of the perturbed instanton density.  The Hamilton-Jacobi
equation implied by (3) equips the moduli space with a rich geometrical
structure which we explore.

\noindent
These results tentatively suggest a picture in which the one-instanton
sector of $SU(2)$ Yang-Mills theory (rather than some large-$N$ limit)
is in some sense holographically dual to bulk gravity.
\end{abstract}
%\end{titlepage}
\newpage
\begin{small}
\tableofcontents
\end{small}

\section{Introduction}

A large amount of work has been done on the AdS/CFT correspondence
\cite{jm,gkp,ewads,adscft}, in particular in the context of supergravity on
$AdS_5\times S^5$ and its dual relationship to ${\cal{N}}=4$ $SU(N)$
super Yang-Mills theories on the boundary of $AdS_{5}$ in the large
$N$ limit.  It was realized early on that instantons of the boundary
gauge theory play a central role in this correspondence \cite{bgkr}.

In particular, in a very interesting series of papers, Dorey et al \cite{dorey}
were able to compute non-perturbative corrections due to Yang-Mills
instantons and show that in the bulk $AdS_5\times S^5$ they correspond
to the contributions of D(-1)-brane instantons in type IIB theory to
the $R^4$ couplings. What is remarkable is that, in the large $N$ saddle
point approximation, the single super-instanton moduli space collapses to
$AdS_5\times S^5$ with $AdS_{5}$ arising as the instanton number $k=1$
$SU(2)$ moduli space.  This work was generalized to other situations
with less supersymmetry. For example, in ${\cal N}=2$ theories arising
in Type I' D3 branes, the Yang-Mills instanton sees the relevant bulk
geometry $AdS_5\times S^5/Z_2$ as well as the geometry of D-7 branes
appearing in the Type I' vacuum \cite{n2}. Related D-instanton probe
calculations appear in \cite{thvkfmt}. 

These results suggest that Yang-Mills instantons are a good probe of the
bulk geometry in general. More precisely, the instanton moduli space in
the large $N$ saddle point limit becomes the bulk geometry.  In particular
the $AdS_5$ coordinates are provided by the position and the scale of the
instanton, with zero scale (UV regime) being the boundary of $AdS_5$ and
large scale (IR regime) being the deep interior. In obtaining this, one
integrates out the zero modes corresponding to the gauge orientation of
the instanton in $SU(N)$ (whose number grows linearly with $N$) leaving
effectively just the moduli associated with $SU(2)$ instantons. The
details of the internal space (such as $S^5$) depend on the R-symmetry
of the problem and hence the number of supersymmetries. These coordinates
appear explicitly as certain bilinears of fermionic zero modes.

In the light of these results it is natural to ask if the feature
that instanton moduli space gives rise to the bulk geometry persists in
situations where Yang-Mills theory on the D3-branes couples to non-trivial
bulk field backgrounds. For example, if the 4-dimensional space-time
on which Yang-Mills theory lives is curved, then does the instanton
moduli space give rise to the deformed $AdS_5$ which is Einstein and
whose metric approaches the 4-dimensional metric as one goes to the
boundary? The AdS/CFT correspondence together with the instanton probe
idea would suggest that this is the case. The purpose of this paper is
to study this problem.

As mentioned earlier, in the large $N$ saddle point effectively
only the moduli associated with $SU(2)$ instantons survive. We will
therefore analyze this problem by directly working with these $SU(2)$
instantons. The first question to settle is what is the metric on the
$k=1$ $SU(2)$ instanton moduli space one should consider. Even in the flat
case this is not {\em a priori} clear as the usual $L^2$-metric on the
moduli space will clearly not give rise to the $AdS_5$ metric. Indeed
the isometries of $AdS_5$ are related to conformal invariance on the
boundary while the $L^2$-metric is not conformally invariant.

Remarkably, there exists a metric on the moduli space, called the
Information Metric, first suggested in the moduli space context in
\cite{hitchin}, which is manifestly gauge and conformally invariant. It
is defined by
\be
G_{AB}(y)
\sim\int\sqrt{g}d^{4}x\; F^{2}(x;y)\del_{A}\log F^{2}(x;y)\del_{B}\log F^{2}(x;y)\;\;.
\label{I}
\ee
Here $\del_{A}$ is the derivative with respect to the instanton moduli
$y^{A}$ and $F^{2}(x;y)$ is short-hand for the instanton density.  It
follows from symmetry arguments \cite{hitchin} or by explicit calculation
(section 2.2) that for the flat metric on $\RR^{4}$ this metric, in fact,
gives the $k=1$ $SU(2)$ instanton moduli space the geometry of $AdS_5$.

The purpose of this paper is to propose a central role for the
Information Metric on instanton moduli space in the context of the AdS/CFT
correspondence.  Specifically, we propose that for an arbitrary boundary
metric the AdS/CFT bulk space-time {\em is} the instanton moduli space
equipped with the Information Metric.

Our aim is thus to study instanton solutions on a curved 4-dimensional
space (which is topologically $\RR^{4}$ or $S^4$) and compute the
corresponding Information Metric on the moduli space. For a general
boundary metric this is a difficult task. So to test the proposal,
we examine the variation of the instanton moduli and the instanton
density $F^{2}$ and the induced variation of the Information metric for
first-order perturbations of the boundary space-time metric away from
the (conformally) flat metric.  We obtain a number of interesting and
technically as well as mathematically quite surprising results, namely
that to this first order in the metric perturbation

\begin{enumerate}
\item the Information metric is Einstein and approaches the perturbed space-time
metric on the boundary - in other words, the variation of the Information Metric
{\em is} the boundary-to-bulk graviton propagator;
\item for this perturbed bulk metric,
the perturbed instanton density $F^2(x;y)$ is the boundary-to-bulk
massless scalar propagator from the point $x$ at the boundary to
the point $y$ in the bulk;
\item the regularized geodesic distance between a point $x$ at the
boundary and an arbitrary point in the
interior labelled by the instanton moduli $y$ is proportional to 
the logarithm of the perturbed instanton action density $F^2(x;y)$.
\end{enumerate}

We also show that result (2) and (3) above imply (1) to first order,
and that to the second order not all the three results above can
simultaneously hold. It would be interesting to see what exactly
happens at the next order, but the techniques used in this paper are
too cumbersome to apply.

The fact that these results are true to first order by itself is quite
remarkable. (1) already shows that the Information Metric is the `correct'
bulk metric to first order in the metric perturbation. And (2) and (3)
show that it is indeed extremely natural to think of the bulk metric
as the Information Metric on the instanton moduli space. It somehow
suggests that self-dual solutions with instanton number $k=1$ of $SU(2)$
Yang-Mills theory are holographically dual to bulk gravity theory, with
the choice of Information metric being dictated by 4-dimensional conformal
invariance which (as in string theory) implies the bulk (target-space)
Einstein equations. We will comment on this scenario in the conclusion.

The paper is organized as follows. In section 2, we briefly review the
Information metric for the unperturbed case. We will see that for the
unperturbed case all the three results stated above are true.  We will
also obtain some useful identites that will be used later in the paper. In
section 3, we discuss the first order correction to the instanton solution
and compute the corresponding correction to the Information metric. We
will show that all the three results hold up to this order. In section
4, we discuss the general structure of the geometry of moduli space
implied by the result (3) or, more precisely, by the Hamilton-Jacobi
equation following from it. We will then explore the relations between
the above three facts and in particular show that result (1) is implied
by (2) and (3). We will also obtain certain harmonic coordinates on the
moduli space which are expressed in terms of the action density and its
$x$-derivatives.  Our results certainly raise more questions than they
answer, so in section 5 we address some of these issues.

\section{The Information Geometry of the Unperturbed Instanton Moduli Space $\M_{0}$}

\subsection{The Information Metric on the Instanton Moduli Space}

The most important property of the Information Metric 
\bea
G_{AB}(y)
&\sim&\int\sqrt{g}d^{4}x\; F^{2}(x;y)\del_{A}\log F^{2}(x;y)\del_{B}\log F^{2}(x;y)\non
&=&\int\sqrt{g}d^{4}x\; \frac{\del_{A} F^{2}(x;y)\del_{B} F^{2}(x;y)}{F^{2}(x;y)}
\label{i3}
\eea
($F^{2}\sim g^{\mu\a}g^{\nu\b}F^{a}_{\mu\nu}F^{a}_{\a\b}$ - 
later on we will fix the precise normalizations we are going to use)
for our purposes is that it inherits all the space-time symmetries of
the instanton density and equations. In particular, it is completely
gauge invariant and hence degenerate along the directions in the moduli
space corresponding to global gauge rotations.  It also depends only on
the conformal structure, i.e.\ the conformal equivalence class of the
space-time metric $g_{\mu\nu}$.  This is manifest from the expression
(\ref{i3}) as both $\sqrt{g}F^{2}$ and the logarithmic derivatives of $F^{2}$
are conformally invariant.

This construction of a metric on the instanton moduli space is a special
case of a general construction of a metric on a space of probability
densities. This {\em Fisher Information Metric} has numerous applications
in statistics and probability theory - for details and references see
e.g.\ \cite{gm,books}.

Clearly (\ref{i3}) is rather different from the standard $L^{2}$-metric
\be
g_{AB}(y)\sim \int\sqrt{g}d^{4}x\;
g^{\mu\nu}(x)\del_{A}A^{a}_{\mu}(x;y)\del_{B}A^{a}_{\nu}(x;y) 
\label{i2}
\ee
on the instanton moduli space which is not conformally invariant but is non-degenerate
in the direction of global gauge moduli.

The properties described above make the Information Metric an extremely
natural object to consider when studying the geometry of moduli spaces.
However, these attractive features are somewhat offset by the fact that
it appears to be quite difficult in general to establish if or when
the Information Metric is non-degenerate (as defined in (\ref{i3}) it
only gives a semi-positive-definite quadratic form) and this may be the
reason why it has received little attention in the differential-geometric
context - the only articles we are aware of are \cite{hitchin,murray,gm}.

Moreover, even if one can prove in certain cases that the Information
Metric is indeed non-degenerate, an explicit or reasonably general closed
form expression for the inverse metric is hard to come by.  For the same
reason, calculations of the curvature of the metric are difficult and
practically nothing seems to be known in cases where the Information
Metric is not known explicitly. Here we will bypass this problem (and
determine the curvature of the Information Metric, among other things)
by working with a perturbation of the Information Metric around a known
background.

\subsection{Explicit Evaluation of the Information Metric on $\M_{0}$}

Let us now consider specifically the Information Metric on the instanton
moduli space $\M_{k=1}(\RR^{4},SU(2))$ of $k=1$ instantons on $\RR^{4}$
with the standard metric. We will denote this space by $\M_{0}$, the subscript
indicating that this is the moduli space associated to the standard
flat metric on $\RR^{4}$ or (by conformal invariance) the standard
round metric on $S^{4}$.

This space is known to be five-dimensional,
with the topology of a ball, thus parametrized by five moduli, namely
four coordinates $a^{\mu}$ (the centre of the instanton) and one scale
$\rho$. As the instanton equations are invariant under conformal $SO(5,1)$
transformations of $\RR^{4}$, the conformal group acts via isometries
on $\M_{0}$ equipped with the Information Metric. Thus, the argument goes
\cite{hitchin}, the Information Metric is the (unique up to a scale)
$SO(5,1)$-invariant metric on the five-ball, i.e.\ the hyperbolic
(Euclidean AdS) metric. Let us check this explicitly.

Explicit expressions for the field strength (in the regular gauge)
and the instanton density as functions of the moduli $y^{A}=(\rho,a^{\mu})$ are 
\bea
F_{\mu\nu}^{a}(x;\rho,a^{\mu}) 
&=& -4\eta^{a}_{\mu\nu} \frac{\rho^{2}}{[(x-a)^{2}+\rho^{2}]^{2}}
\\
\tr F^{2}(x;\rho,a^{\mu}) 
&=& 96\frac{\rho^{4}}{[(x-a)^{2}+\rho^{2}]^{4}}\;\;.
\label{trf2}
\eea
Here $\eta^{a}_{\mu\nu}$ are the 't Hooft eta-symbols, a basis for
self-dual two-forms on $\RR^{4}$ - see (\ref{sigmai1}).

Let us define
\be
F^{2} := \frac{\rho^{4}}{[(x-a)^{2}+\rho^{2}]^{4}}\;\;.
\label{i4}
\ee
Since
\be
\frac{6}{\pi^{2}} \int d^{4}x\; F^{2} = 1
\ee
we define the Information Metric by
\be
G_{AB} = c \frac{6}{\pi^{2}}\int d^{4}x\;F^{2}\del_{A}\log F^{2}\del_{B}\log F^{2} 
\;\;,
\ee
where $c$ is a constant which we will choose in a convenient way below.

The logarithmic derivatives $\del_{A}\log F^{2}$ 
of $F^{2}$ with respect to the moduli are
\bea
\del_{\rho}\log F^{2} 
&=& 4\left[\frac{1}{\rho}-\frac{2\rho}{(x-a)^{2}+\rho^{2}}\right]\non
\del_{a^{\mu}}\log F^{2}
&=& 8\frac{(x-a)_{\mu}}{(x-a)^{2}+\rho^{2}}\;\;.
\label{i5}
\eea
One then finds
\be
G_{a^{\mu}a^{\nu}} = c\frac{16}{5}\frac{\d_{\mu\nu}}{\rho^{2}}\;\;,
\;\;\;\;\;\;
G_{\rho\rho} =c\frac{16}{5} \frac{1}{\rho^{2}}\;\;,\;\;\;\;\;\;
G_{a^{\mu}\rho} = 0\;\;.
\ee
This establishes that the Information Metric on $\M_{0}$ is
\be
ds_{0}^{2} \equiv G_{AB}dy^{A}dy^{B} = 
\frac{16c}{5}\frac{d\rho^{2} + d\vec{a}^{2}}{\rho^{2}}\;\;.
\ee
Choosing $c=5/16$, i.e.
\be
G_{AB}= \frac{15}{8\pi^{2}}\int d^{4}x\;F^{2}\del_{A}\log F^{2}\del_{B}\log F^{2}\;\;,
\label{i6}
\ee
we thus get precisely the unit hyperbolic metric. This is
a smooth conformally invariant and geodesically complete metric on $\M_{0}$. We will
adopt this normalization for the Information Metric in general.

By contrast, the $L^{2}$-metric on $\M_{0}$ for $\RR^{4}$, the flat metric
\be
ds^{2} = d\rho^{2}+  d\vec{a}^{2}
\ee
with a singularity at $\rho=0$ corresponding to zero-size instantons,
is neither conformally invariant nor geodesically complete.  Also the
$L^{2}$-metric for $S^{4}$ is quite different - and somewhat more
complicated. It has been studied in \cite{gplh1}. Note that the
Information Metric, on the other hand, is of course the same for $\RR^{4}$
and $S^{4}$ because of conformal invariance!

Let us also note that the Information Metric on the moduli 
space $\M_{k=1}(\RR^{4},SU(N))$ of $k=1$ $SU(N)$-instantons
collapses to that on
\be
\M_{0}=\M_{k=1}(\RR^{4},SU(2))\subset\M_{k=1}(\RR^{4},SU(N))\;\;.
\ee
Indeed, for $k=1$ and gauge group $SU(N)$ one can construct all instanton
solutions by embedding the $SU(2)$-instanton inside $SU(N)$ and acting
with rigid $SU(N)$ gauge transformations. 
The $L^{2}$-metric on the resulting $4N$-dimensional moduli 
space $\M_{k=1}(\RR^{4},SU(N))$ is non-degenerate.  

The Information Metric, on the other hand, is, as should be clear from
what has been said above, very different. In fact, since the instanton
density is invariant under all gauge transformations, the Information Metric 
on $\M_{k=1}(\RR^{4},SU(N))$ is completely degenerate along
the $SU(N)$ gauge moduli directions and thus reduces to the $k=1$
$SU(2)$ Information Metric on $\M_{k=1}(\RR^{4},SU(2))$, i.e. the
$AdS_{5}$-metric, for any $N$.

While this is an intriguing result, in particular in light of the AdS/CFT
emergence of $AdS_{5}$ in the large-$N$ instanton calculus \cite{dorey},
we do not quite know what to make of this. 

%We will come back to this in the Conclusions.

\subsection{The Groisser-Murray Theorem and the Relation to the Fefferman-Graham
Construction}

As shown above, it follows from the conformal invariance of the
Information Metric that on $\M_{k=1}(S^{4},SU(2))$ it coincides (up
to a scale) with the standard hyperbolic (Euclidean AdS) metric on
the five-ball.

Groisser and Murray have shown \cite{gm} that this is prototypical for the
asymptotic behaviour of $G_{AB}$ on the $k=1$ $SU(2)$ moduli space for a
much larger class of manifolds, in particular for $S^{4}$ (or $\RR^{4}$)
equipped with a  metric which is not the standard (round or flat) metric.

Intuitively one would expect nearly point-like instantons on such a space
to probe the geometry of the space-time $X$ and its metric $g_{X}$.
And the main result of Groisser and Murray is that indeed, as a suitably defined
scale function $\rho\ra 0$, the Information Metric $G_{AB}$ on the 5-dimensional
moduli space $\M$ tends to
\be
ds^{2} \sim \frac{d\rho^{2} + g_{X}}{\rho^{2}}\;\;.
\label{gm}
\ee
In particular, therefore, for sufficiently small $\rho$ the Information
Metric is non-degenerate and asymptotically hyperbolic.

The encouraging thing about this result is that this is precisely the
asymptotic form of the metric entering the Fefferman-Graham construction
\cite{fg} and expected from the AdS/CFT correspondence. 

Recall that in the Fefferman-Graham construction one makes an ansatz
for the bulk metric $G_{AB}$ which is to satisfy the Einstein equations
with a negative cosmological constant, and with boundary value $g_{X}$,
of the form
\be
ds^{2} = \frac{d\rho^{2} + g_{X}(\rho)}{\rho^{2}}\;\;,
\ee
where 
\be
g_{X}(\rho) = g_{X} + \rho^{2} g_{X}^{(2)} + \rho^{4}g_{X}^{(4)} + \ldots
\;\;.
\ee
For $\rho\ra 0$ this agrees precisely with the $\rho\ra 0$ behaviour 
of the Information Metric. This means that the asymptotic behaviour of 
the metric is compatible with the bulk Information Metric being Einstein.

Note that, even though we know that the Information Metric is conformally
invariant, the asymptotic form of the Information Metric given above
appears to depend on the metric $g_{X}$ itself, not only on its conformal
class $[g_{X}]$.

However, here yet another useful fact about asymptotically AdS
spaces comes to the rescue, namely that Weyl transformations of 
the boundary metric can be induced by certain bulk diffeomorphisms.
These have been discussed in detail in \cite{isty}. 
Thus Weyl invariance of the Information Metric is indeed compatible with
the Groisser-Murray form of the asymptotic metric, as the Information
Metrics for different conformal factors are all diffeomorphic.

Moreover, the analysis of \cite{isty} shows that the first non-trivial
term $g_{X}^{(2)}$ in the Fefferman-Graham expansion is determined
uniquely by conformal invariance alone and agrees with the result
obtained by Henningson and Skenderis \cite{mhks} by solving the bulk
Einstein equations to that order.

This means that, were we to calculate the Information Metric in a
small-$\rho$ expansion (we will not), then to first non-trivial order the
result would be guaranteed to be compatible with the Information Metric
being Einstein. At higher orders, however, conformal invariance does not
fix the coefficients $g_{X}^{(2n)}$ uniquely as there are non-trivial
higher-derivative Weyl invariants like the Euler density and the square
of the Weyl tensor.

\subsection{Other Aspects of the Information Geometry of $\M_{0}$}

In this section we will analyse and summarise some aspects of the geometry
of $\M_{0}$ equipped with the Information Metric. Of course the geometry
of the five-dimensional hyperbolic plane or Euclidean AdS space is well
known, and it is not our intention to review these facts here.

Rather, the purpose of this section is to focus on and highlight those
aspects of the AdS/information geometry which we will show survive to
first order in the metric perturbation. In particular, therefore, the
(super-)symmetries of AdS will play no role in the following. In passing
we will also list some identities which will be useful for
the calculations of section 3.

One other aspect of the geometry of AdS, namely the boundary-to-bulk
geodesic distance functional, which will turn out to play a fundamental
role in the following, will be dealt with seperately in the next section.

Thus we consider the standard AdS metric 
\be
ds_{0}^{2} \equiv G_{AB}dy^{A}dy^{B} = \frac{d\rho^{2} + d\vec{a}^{2}}{\rho^{2}}\;\;,
\label{b1}
\ee
which, as we know, has the integral representation
\be
G_{AB}= \frac{15}{8\pi^{2}}\int d^{4}x\;F^{2}\del_{A}\log F^{2}\del_{B}\log F^{2}\;\;,
\ee
where $F^{2}$ 
was defined in (\ref{i4}).
This metric is maximally symmetric, i.e.\ its curvature tensor is
\be
R_{ABCD} = -(G_{AC}G_{BD}-G_{AD}G_{BC})
\ee
and, in particular, Einstein,
\be
R_{AB} = - 4 G_{AB}\;\;.
\label{b3}
\ee
The function (actually four-parameter family of functions) $F^{2}(x;\rho,a^{\mu})$
on $\M_{0}$
satisfies two interesting identities.  Let $\nabla_{A}$ denote the
AdS covariant derivative and $\Box$ the scalar Laplacian,
\be
\Box = G^{AB}\nabla_{A}\nabla_{B} = \rho^{2}[\del_{\rho}\del_{\rho}
+\d^{\mu\nu}\del_{a^{\mu}}\del_{a^{\nu}}-3 \rho^{-1}\del_{\rho}]\;\;.
\label{b4}
\ee
Then the first identity is
\be
\Box F^{2} = 0\;\;.
\label{b5}
\ee
As
\be
\lim_{\rho\ra 0}\frac{6}{\pi^{2}}F^{2}(x;\rho,a^{\mu}) = \d(a-x)\;\;,
\ee
this establishes that $(6/\pi^{2}) F^{2}$ is the {\em boundary-to-bulk propagator}
of a massless scalar field on $AdS_{5}$ \cite{ewads}. I.e.\ for an arbitrary
boundary scalar field $\phi(x)$, the bulk scalar $\Phi(\rho,a^{\mu})$ defined by
\be
\Phi(\rho,a^{\mu})= \frac{6}{\pi^{2}}\int d^{4}x\;F^{2}(x;\rho,a^{\mu})\phi(x)
\ee
satisfies
\be
\Box \Phi(\rho,a^{\mu}) = 0
\ee
and
\be
\lim_{\rho\ra 0}\Phi(\rho,a^{\mu})=\phi(a^{\mu})\;\;.
\ee
It is also true that the massive scalar propagator is given by some power of
$F^{2}$. Indeed it can be verified that
\be
\Box (F^{2})^{\Delta/4} = \Delta(\Delta-4)(F^{2})^{\Delta/4}\;\;.
\label{b6}
\ee
In particular, $(F^{2})^{1/2}$ corresponds to the tachyonic propagator with mass
$m^{2}=-4$. The  identity (\ref{b6}) for all $\Delta$
follows from (\ref{b5}) and the rather remarkable identity
\be
G^{AB}(y)\del_{A}\log F^{2}(x;y)\del_{B}\log F^{2}(x;y) =
16\;\;\;\;\;\;\forall\;\;x,y\;\;, 
\label{b8}
\ee
which can be readily verified by explicit calculation using the expressions for 
the logarithmic derivatives of $F^{2}$ given in (\ref{i5}). 

These logarithmic derivatives of $F^{2}$ will be ubiquitous in the following,
and it will be convenient ot think of them as a four-parameter family (indexed
by $x$) of covector fields on $\M_{0}$ we will call (with a convenient normalization)
\be
\va(x):=\trac{1}{4}\del_{A}\log F^{2}\;\;.
\label{vax}
\ee
Obviously these vector fields satisfy
\be
G^{AB}\va(x)\vb(x) = 1\;\;\;\;\;\;\forall\;\;x
\label{b9}
\ee
and
\be
\nabla_{A}\vb =\nabla_{B}\va\label{b13}
\ee
Some of their other main properties are 
\be
\nabla_{A}\vb = - G_{AB} + \va\vb\label{b10} 
\ee
and
\bea
v^{A}\nabla_{B}\va &=& 0 \label{b14}\\
v^{A}\nabla_{A}\vb&=& 0\label{b15}\\
G^{AB}\nabla_{A}\vb &=& -4\;\;.\label{b11}
\eea
Property (\ref{b14}) follows from 
(\ref{b9}) by differentiation; (\ref{b15}) says that $\va$ is
a geodesic vector field and follows from (\ref{b13}, \ref{b14});
lastly, given (\ref{b8}), (\ref{b11}) is equivalent to  (\ref{b5}).

Geometrically (\ref{b9}) means that we have a four-parameter family
of constant norm or geodesic gradient vector fields on $\M_{0}$. It
is easy to see that these span the tangent space at each point of the
moduli space.

In the following we will refer to (\ref{b5}) and (\ref{b9}) as the
Propagator Equation and the Hamilton-Jacobi (HJ) Equation respectively
(the reason for calling (\ref{b9}) the HJ equation will emerge below).
We will show that these equations, as well as the Einstein equation,
continue to be valid to first order in the metric perturbation, and we
will study the resulting geometry imposed on $\M_{0}$ and its perturbation
$\M$ in much greater detail in section 4. In particular, the power
of these identities derives form the fact that they hold for all $x$.
Thus new identities on $\M$ can be derived from them by differentiation
with respect to $x$.

Property (\ref{b10}), on the other hand, which immediately implies
(\ref{b14})-(\ref{b11}), also implies that the metric $G_{AB}$
is maximally symmetric and can thus not hold in general. Indeed,
differentiating once more and taking commutators, one finds
\be
[\nabla_{C},\nabla_{A}]\vb = (G_{AB}G_{CD}-G_{BC}G_{AD})v^{D}\;\;.
\ee
Since this holds for all $x$, and the $\va(x)$ span the tangent space of $\M_{0}$,
this implies that the curvature tensor of the bulk Information Metric is 
\be
R_{BDAC} = -(G_{AB}G_{CD}-G_{BC}G_{AD})\;\;.
\label{a18}
\ee

The identity (\ref{b10}) has one unobvious consequence
that will play a role in the following, namely that
\be
\int d^{4}x\; F^{2}(x) \va(x)\vb(x)\vc(x) =0 \;\;\;\;\;\;\forall\;\; A,B,C\;\;.
\label{b12}
\ee
To prove this, we use the fact that the Information Metric can be 
written in integral form as
\be
G_{AB}\sim \int F^{2} \va\vb\;\;.
\ee
Since by definition 
\be
\nabla_{C}G_{AB}=0\;\;,
\ee
it follows that
\be
\nabla_{C}\int F^{2} \va\vb = 0\;\;.
\ee
Using
\be
\int F^{2}\va = \trac{1}{4}\int \del_{A}F^{2} =\trac{1}{4} \del_{A}\int F^{2} = 0
\ee
and (\ref{b10}), (\ref{b12}) follows.

One more identity that we will need is the `two-point' version of 
(\ref{b9}). Namely, since $v^{A}(x)\va(x)=1$, it must be true that 
$v^{A}(x)\va(y)-1$ is proportional to $|x-y|$. The precise formula is
\be
v^{A}(x)\va(y)-1=\frac{2\rho^{2}(x-y)^{2}}{[(x-a)^{2}+\rho^{2}][(y-a)^{2}
+\rho^{2}]}\;\;.
\label{b16}
\ee

\subsection{The Boundary-to-Bulk Geodesic Distance in AdS and the Instanton Density}

The aim of this section is to show that 
\be
-\trac{1}{4}\log F^{2} =\log \frac{(a-x)^{2} + \rho^{2}}{\rho}
\ee
can be interpreted as the regularized geodesic distance in AdS between the
bulk point $(\rho,a)$ and the boundary point $(0,x)$ (regularized because
the geodesic distance to the boundary at $\rho=0$ is infinite). This
provides an explanation for the validity of (\ref{b9}) as it can now be
interpreted as the Hamilton-Jacobi equation for the classical geodesic
action functional, as we will explain in section 2.6.

There are a number of different ways to establish this result. One is to integrate
the geodesic equation directly. Alternatively one can
use the {\em chordal distance} $L$ of $AdS_{d+1}$, i.e.\ the restriction to
\be
\eta_{AB}X^{A}X^{B} \equiv (X^{1})^{2} + 
\ldots + (X^{d+1})^{2} - (X^{0})^{2} = -1\;\;.
\ee
of the unique $SO(d+1,1)$-invariant distance 
\be
L(X_{0},X_{1}) := \eta_{AB}(X_{1}^{A}-X_{0}^{A})(X_{1}^{B}-X_{0}^{B})
\ee
on the embedding space. A straightforward calculation shows that in Poincar\'e
coordinates $(\rho,a^{\mu})$ one has
\be
L(\rho_{0},a_{0};\rho_{1},a_{1})=
\frac{(\rho_{1}-\rho_{0})^{2}+(a_{1}-a_{0})^{2}}{\rho_{0}\rho_{1}}\;\;.
\ee
Because of the isometries of $AdS_{d+1}$, the geodesic distance function $D$
can only be a universal function of the chordal distance $L$. Thus to
determine this universal function, it is sufficient to consider purely radial
geodesics ($a_{1}=a_{0}$) for which evidently the geodesic distance
is 
\be
D(\rho_{0},a;\rho_{1},a)=\log\frac{\rho_{1}}{\rho_{0}}\;\;.
\ee
This is related to the chordal distance
\be
L(\rho_{0},a;\rho_{1},a)=
\frac{\rho_{1}}{\rho_{0}} + \frac{\rho_{0}}{\rho_{1}}-2
\ee
by
\be
L+2=\ex{D} + \ex{-D}=2\cosh D
\ee
or
\be
L = 4 \sinh^{2}(D/2)\;\;,
\ee
which is thus the general relation between the chordal and geodesic distance.
For $L$ very large, the relation becomes
\be
D\approx\log L\;\;.
\ee
In particular, therefore,
\bea
\lim_{\epsilon\ra 0} D(\epsilon,x;\rho,a) &=& 
\log\frac{\rho^{2} + (a-x)^{2}}{\rho\epsilon}\non
&=& \log\frac{\rho^{2} + (a-x)^{2}}{\rho} -\log\epsilon\;\;,
\label{gd}
\eea
as was to be shown.

While this is a quick and clean argument, it has the 
disadvantage of relying on the isometries of AdS, and can therefore not generalize
to the perturbed instanton moduli spaces we are interested in. We will now give an
argument that, as we will see later, does generalize since it only relies on the
validity of the HJ equation (\ref{b9}).

We know from (\ref{b15}) that the $\va$ are geodesic vector fields,
\be
v^{A}\nabla_{A}\vb = 0\;\;.
\label{gd1}
\ee
Let $y^{A}(\tau)=(\rho(\tau),a(\tau))$ be the corresponding geodesic curves. 
Thus we have
\be
\dot{y}^{A} = G^{AB}\vb\;\;.
\ee 
To integrate this equation, we calculate
\be
\frac{d}{d\tau}\log F^{2} = \dot{y}^{A}\del_{A}\log F^{2} = 4 v^{A}\va = 4\;\;,
\ee
so that these geodesic curves can be simply described by the equation
\be
\log F^{2}(x;y(\tau)) = \log F^{2}(x;y(\tau=0)) + 4\tau\;\;.
\ee
Thus the proper distance is 
\be
D=\int_{\tau=0}^{\tau=\tau_{+}}d\tau = 
\trac{1}{4}\log F^{2}(x;y(\tau_{+}))-\trac{1}{4} \log F^{2}(x;y(\tau=0))\;\;.
\ee
We choose $y^{A}(0) = (a_{0},\rho_{0})$ and $y^{A}(\tau_{+}) = (\epsilon,a_{+})$ and
note that for $\epsilon\ra 0$ we necessarily have $\tau\ra\infty$. It remains to
show that as $\epsilon\ra 0$ we also have $a_{+}=x$. This follows from the
fact that
\be
\lim_{\epsilon\ra 0}\log F^{2}(x;\epsilon,a) = \pm \infty\;\;,
\label{gd2}
\ee
the upper sign occuring if $x=a$, and the lower if $x \neq a$. As the geodesic 
distance is a positive quantity (the $v^{A}$ point towards the boundary and we are
integrating towards the boundary), we must have the upper sign and hence $x=a_{+}$.
Regularizing the geodesic distance as before by removing this infinity, we reproduce
precisely the result (\ref{gd}).

\subsection{$G^{AB}\va\vb=1$ is the Hamilton-Jacobi Equation for Geodesic Motion on
$\M_{0}$}

Since we have shown that 
$D= -\trac{1}{4}\log F^{2} -\log\epsilon$ 
is the geodesic distance,
it must satisfy the Hamilton-Jacobi equation (Hamiltonian constraint)
\be
G^{AB}p_{A}p_{B}=1\;\;,
\ee
where the canonical momentum $p_{A}$ is obtained by varying the classical
geodesic action $S_{cl}(y)=D(\epsilon,x;y)$ with respect to the end-point
$y=(\rho,a)$,
\be
p_{A}=\del_{A}D(\epsilon,x;y)
\ee
In particular, the $p_{A}$ are finite as $\epsilon\ra 0$. Thus the Hamilton-Jacobi
equation for boundary-to-bulk paths is
\be
G^{AB}\del_{A}D(0,x;y) \del_{B}D(0,x;y)=1\;\;,
\ee
which is nothing other than the equation $G^{AB}\va(x)\vb(x)=1$.

We will show later that the relation between the geodesic distance and the instanton
density continues to hold to first order in the metric perturbation, and this may
provide a rationale for the validity of the (in other respects somewhat mysterious)
HJ equation to that order.

Before leaving this topic we would like to point out that these results
suggest that a WKB type approximation to the path integral for the
boundary-to-bulk scalar propagator should be exact.

\section{First Order Corrections}

In this section we will determine the first order corrections to the
Information metric, massless scalar boundary-to-bulk propagator
and the geodesic distance (Hamilton Jacobi equation) when one deforms
the 4-dimensional space on which the Yang-Mills theory lives. The
first step in this computation therefore is to calculate the 
correction to the instanton solution to first order in the 
perturbation of the boundary space-time metric. 

\subsection{Metric Deformations and Instanton Deformations}

To determine the change in the Information Metric on the instanton
moduli space as we vary the metric on $S^{4}$ requires knowledge of
the variation of the instanton density 
\be
\tr F_{A}*F_{A}=\tr F_{A}\wedge F_{A}\;\;,
\ee
i.e.
\be
\delta_{h} \tr (F_{A}\wedge F_{A}) = 2 \tr ( F_{A} \wedge\delta_{h} F_{A}) 
\label{dense}
\ee
where $\delta_{h}$ represents the variation $\d_{h}g_{\mu\nu}=h_{\mu\nu}$
of the boundary metric. Notice that since $F_{A}$ is
self-dual only the self-dual part of the variation $\delta_{h} F_{A}$
enters on the right hand side. However, the deformation of the self-duality 
equations gives an algebraic equation only for the anti-self-dual 
part of $\delta_{h} F_{A}$, see (\ref{om}) below. This means that we
need to get an explicit formula for the deformed instanton so as to be
able to determine (\ref{dense}).  

The ideas here go back to the work of Taubes \cite{taubes} on the
`grafting' of instantons from $S^{4}$ to an arbitrary 4-manifold. In
our case we are grafting the instantons on $S^{4}$ to an $S^{4}$ which
is equipped with a different metric. Let $*_{0}$ be the Hodge star
operator for the round metric on $S^{4}$ and let $* =*_{0} + \delta_{h} *
$ denote the perturbed Hodge operator. $\delta_{h} * = t*_{1}+ t^{2}*_{2}
+ \dots$. Here we work to first order in $t$. In local coordinates we
write the metric as $g_{\mu \nu} = g^{0}_{\mu \nu} + t \, h_{\mu \nu }
+ \dots$. While the above refers to $S^{4}$ all of our formulae are
written in the context of ${\Bbb R}^{4}$.

We wish to solve the equation  
\be 
F_{A} = * F_{A}.
\label{inst} 
\ee
Write the connection as 
\be 
A = A_{0} + t\omega + \dots 
\label{omega}
\ee 
where $A_{0}$ is the instanton solution on the round $S^{4}$. Differentiating
(\ref{inst}) once with respect to $t$ and setting $t=0$ we obtain 
\be
(1-*_{0})d_{A_{0}}\omega = *_{1} \, F_{A_{0}}, 
\label{om} 
\ee
so that, in particular, $*_{1}F_{A_{0}}$ is anti-self-dual.

The choice of $\omega$ is dictated by the requirements that it is
perpendicular to gauge transformations and also does not represent a
tangent vector to the moduli space $\cal{M}_{0}$. One takes  
\be 
\omega = *_{0}d_{A_{0}} u
\label{u}
\ee 
where $u=\frac{1}{2}u_{\mu\nu}dx^{\mu}dx^{\nu}$ is an anti-self-dual two-form. 
With this choice (\ref{om}) becomes
an elliptic equation,  
\be 
\Delta_{0} u = *_{1} F_{A_{0}}\;\;, 
\label{lap}
\ee 
where 
\be 
\Delta_{0} = - \left(d_{A_{0}}*_{0}d_{A_{0}}*_{0}+*_{0}
d_{A_{0}}*_{0} d_{A_{0}}\right)
\ee
is the Laplacian on two-forms, so that one can solve for $u$. 
In local coordinates, (\ref{u}) is
\be
\omega_{\mu} =-D_{\la}^{0}\omega^{\la}_{\;\mu}\;\;,
\ee
and the Green's function form of (\ref{lap}) reads
\be
D_{0}^{2} u_{\mu \nu}(x,y) = -\delta(x-y) \, (*_{1}F_{A_{0}})_{\mu \nu}(y) ,
\ee
where
\be
(*_{1}F_{A_{0}})_{\mu \nu} = h_{\mu}^{\; \alpha} F_{\alpha \nu}(A_{0}) -
h_{\nu}^{\; \alpha}F_{\alpha \mu}(A_{0}).
\ee
Here labels are raised with the original unperturbed metric. Notice also that
there is no term $h_{\alpha}^{\; \alpha}F_{\mu \nu}(A_{0})$
as the instanton equation is conformally invariant. For this reason we will
henceforth only consider traceless perturbations $h_{\mu\nu}$. 
Also, as there is no likelyhood of confusion we will from now on, and until
section 4, drop the zero subscript. 

One can solve
the Green's function equation \cite{bcclcfgt}, with the result that
\be
u_{\mu \nu}(x,y) = \frac{\overline{U}(x)U(y) (*_{1}F)_{\mu \nu}(y) \,
\overline{U}(y) U(x)}{4\pi^2 (x-y)^{2}},
\label{umunu}
\ee
where $U$ is a $4\times 2$ matrix in terms of which the self-dual
gauge potential (at zero'th order) is simply $A_{\mu}= i\bar{U}
\partial_{\mu} U$. Explicitly $U$ is
\be
U(x) = \frac{1}{\sqrt{ (x-a)^{2} + \rho^{2} }} \, \left(
\begin{array}{c}
(x-a)^{\mu} \overline{\sigma}_{\mu}\\
- \rho
\end{array}\right) g(x)
\label{U}
\ee
and $g(x)$ is an unspecified $SU(2)$ gauge group element. Since we
will be dealing with gauge invariant quantities there is no reason to
fix on a particular $g(x)$. 

The $\sigma$-matrices are defined by 
\be
\sigma_{\mu} = (\II,i\tau_{a})\;\;,\;\;\;\;\;\;
\bar{\sigma}_{\mu} = (\II,-i\tau_{a})\;\;,
\ee
where the $\tau_{a}$ are the standard Pauli matrices. They satisfy 
\be
\sigma_{\mu} \bar{\sigma}_{\nu} +\sigma_{\nu} \bar{\sigma}_{\mu}=
2 \delta_{\mu\nu}
\label{sigmai0}
\ee
and are such that 
\be
2\sigma_{\mu \nu} \equiv 
\sigma_{\mu}\bar{\sigma}_{\nu}
-\sigma_{\nu}\bar{\sigma}_{\mu} = 2i \eta^{a}_{\mu\nu}\tau_{a}
\label{sigmai1}
\ee
is self-dual. One has the identities 
\be
\sigma_{\mu}^{~ \lambda} A^{(-)}_{\nu\lambda} - 
\sigma_{\nu}^{~ \lambda} A^{(-)}_{\mu\lambda} = 0\;\;,\;\;\;\;\;\;
\sigma_{\mu}^{~ \lambda} A^{(+)}_{\nu\lambda}+
\sigma_{\nu}^{~ \lambda} A^{(+)}_{\mu\lambda}=\frac{1}{4}
\delta_{\mu \nu} \sigma^{\alpha \beta} A^{(+)}_{\alpha \beta}\;\;, 
\label{sigmai2}
\ee
for any anti-self-dual and self-dual tensors $A^{(-)}$ and $A^{(+)}$
respectively. 

Another useful fact that is often used in computations is that to first order
\bea
\delta_{h} \tr F_{\mu \nu} g^{\mu \alpha} g^{\nu \beta} F_{\alpha \beta} &
= & 4 \tr F_{\mu \nu}\, \delta^{\mu \alpha}\, \delta^{\nu \beta}\,  D_{\alpha}
\, \omega_{\beta} \nonumber \\
& = & 4 \partial^{\mu} \tr F_{\mu \nu}\, \omega^{\nu} \nonumber \\
& = & 4\partial^{\mu} \partial^{\alpha} \tr F_{\mu \nu}\, u^{\nu}_{\; \alpha}\;\;.
\label{dmudnu}
\eea
The first equality follows since if we change the metric the inner
product $\tr F_{A} *_{1} F_{A} = 0$, the second follows by the Bianchi
identity (this is the usual relationship $\delta_{h} \tr F^{2}_{A} = 2d \tr F_{A}
\delta_{h} A$), the third by the Bianchi identity again.

\subsection{The Linearized Bulk Einstein Equation}

We will now compute the first order deformation of the Information metric
(\ref{i6}) arising from the deformation of the flat metric on $\RR^4$ 
via the first order correction to the instanton solution.

{}From the definition of the Information metric it follows that
\be
\delta_{h} G_{AB} = \delta_{h} G'_{AB}  + \nabla_{(A} V_{B)}\;\;,
\label{dgab}
\ee
where
\be 
\delta_{h} G'_{AB} = -\frac{5}{64\pi^2}\int d^4 x\; (\delta_{h} \tr F^2) 
\frac{\nabla_A \nabla_B
\sqrt{\tr F^2}}{\sqrt{\tr F^2}}
\ee
and
\be
V_A = \frac{5}{64\pi^2}\int d^4 x\; (\delta_{h} \tr F^2) \partial_A \ln \tr F^2
\;\;.
\ee
Here we have used the fact that $F^2$ as defined in (\ref{i4}) is
normalized as $1/96$ times $\tr F^2$  defined in (\ref{trf2}). The
covariant derivatives $\nabla_A$ are with respect to the unperturbed
$AdS_{5}$ metric. $\delta_{h} G$ and $\delta_{h} G'$ are therefore
related by a bulk diffeomorphism. 

We will see below that, somewhat unexpectedly but very conveniently, $\delta_{h}
G'_{AB}$ satisfies the transverse-traceless gauge condition. Therefore
the linearized Einstein equation is simply 
\be
\Box \delta_{h} G'_{AB} = (\Lambda/2) \delta_{h} G'_{AB}\;\;,
\label{leeq}
\ee
when the unperturbed metric satisfies $R_{AB}=\Lambda G_{AB}$.

$\delta_{h} G'_{AB}$ can be simplified further by using (\ref{vax}) and
(\ref{b10}) to
\be
\delta_{h} G'_{AB} = -\frac{15}{32\pi^2}\int d^4 x\; v_A v_B \delta_{h} \tr F^2
\;\;,
\ee
where we have used the fact that $\int d^4 x\; \delta_{h} \tr F^2 =0$.

Using $v^A v_A=1$ it follows that $\delta_{h} G'_{AB}$ is traceless,
$G^{AB}\delta_{h} G'_{AB}=0$.
Furthermore, using (\ref{b15}) and (\ref{b11}) one finds 
\bea
\nabla^A \delta_{h} G'_{AB} &=& -\frac{15}{32\pi^2} \int d^4 x\; v_B 
(v^A \nabla_A -4)\delta_{h} \tr F^2 \non
&=& 
-\frac{15}{8\pi^2} \int d^4 x\; F^{2} v_B v^{A}\delta_{h}\va
\;\;.
\label{ddg'ab}
\eea
In the next subsection we will establish that the HJ equation holds to first order
in the metric variation, i.e.\ that
\be
2v^{A}(x)\delta_{h}\va(x) = -\va(x)\vc(x) \d_{h}G^{AC}\;\;.
\ee
Substituting this in (\ref{ddg'ab}) and using (\ref{b12}), it follows that
$\delta_{h} G'_{AB}$ is also transverse.

To proceed further we need the expression for $\delta_{h} \tr 
F^2$. Substituting the explicit expressions for $u_{\mu \nu}$ as
given by (\ref{umunu}) and (\ref{U}) in (\ref{dmudnu}) we have
\be
\delta_{h} \tr F^2(x) = \partial_{\mu} \partial_{\nu} S_{\mu \nu}
\label{df1}
\ee
with
\be
S_{\mu \nu}= -\frac{4}{\pi^2\rho^2}
\int d^4 y (F^2(x) F^2(y))^{\frac{3}{4}} 
\frac{1}{(x-y)^2} \tr (\rho^2 + Y \bar{X})\sigma_{\mu \la} (\rho^2 + X
\bar{Y}) H_{\la\nu} 
\;\;.
\label{df2}
\ee
Here  capital $X$ and $Y$ denote $(x-a)$ and $(y-a)$ respectively,
$\bar{X} = X^{\mu}\bar{\sigma}_{\mu}$, 
$a$ and $\rho$ are the position and scale of the unperturbed
instanton, and
\be
H_{\la\nu}(y) = h_{\lambda}^{~\mu}(y) \sigma_{\mu \nu}- 
h_{\nu}^{~\mu}(y)
\sigma_{\mu \la}
\ee
is anti-self-dual. We have set $h_{\mu}^{\mu}=0$ since this does not 
change the Information Metric due to the conformal invariance. 

We sketch below the computation for $\delta_{h} G'_{\rho \rho}$. The
remaining components of the metric deformation as well as the 
diffeomorphism vector field $V_A$ can be computed in a similar way.

Making use of the $\sigma$-matrix identities and the fact that
$v_{\rho} = (X^2-\rho^2)/\rho(X^2 +\rho^2)$, we find
that
\be
\delta_{h} G'_{\rho \rho} = -\frac{240}{\pi^4} \int d^4 y \int d^4 x 
\frac{\rho^4 (X^2- 2\rho^2)(\rho^2 X^{\mu}+ X^2 Y^{\mu})
(\rho^2 X^{\nu}+ X^2 Y^{\nu})}
{(Y^2 +\rho^2)^3 (X^2 + \rho^2)^7(x-y)^{2}} h_{\mu\nu}(y).
\ee
By writing $X^2=(X^2+\rho^2)-\rho^2$ in the  numerator we can bring the
$x$ integral in the form of sums of terms like $f(X)/(X-Y)^2(X^2
+\rho^2)^n$ where $f(X)$ is a quadratic polynomial of the form
$B_{\mu\nu}X^{\mu}X^{\nu} + B_{\mu}X^{\mu}+B$. 
The $x$-integral of such terms can be performed by using Feynman 
parametrization
\be
\frac{1}{(X^2+\rho^2)^n (X-Y)^2} = \int_1^{\infty} dt \frac{n
(t-1)^{n-1}
t^{n+1}}
{\bigl{[} (tX-Y)^2 + (t-1)(Y^2 + t\rho^2) \bigr{]}^{n+1}}\;\;,
\ee
shifting $X \rightarrow (X  +Y)/t$, and using the formula
\be
\int d^4 X \frac{1}{(X^2 + A)^n} =
\frac{\pi^2}{(n-1)(n-2)}\frac{1}{A^{n-2}} ~~.
\ee
Finally the $t$-integral is easily computed with the result
\be
\delta_{h} G'_{\rho \rho}= \frac{40}{\pi^2}\int d^4 y\frac{\rho^4}
{(Y^2+\rho^2)^6} Y^{\mu} Y^{\nu} h_{\mu \nu}(y)       
\;\;.
\label{g'rhorho}
\ee
Similarly we can compute the other components of the metric
deformation in the transverse traceless gauge and the result is
\begin{eqnarray}
\delta_{h} G'_{\rho a} &=&\frac{20}{\pi^2}\int d^4 y \frac{\rho^3}{(Y^2+\rho^2)^5}
\bigl{[} Y^{\mu} h_{\mu a}(y) - 2 \frac{1}{Y^2+\rho^2} Y^{\mu}
Y^{\nu} h_{\mu \nu}(y) Y_a \bigr{]} \nonumber \\
\delta_{h} G'_{ab} &=& \frac{10}{\pi^2}\int d^4 y  \frac{\rho^2}{(Y^2+\rho^2)^4}
\bigl{[}  h_{ab}(y) - 2 \frac{1}{Y^2+\rho^2}Y^{\mu} h_{\mu (a}(y) Y_{b)} +
\nonumber \\ &~&~~~~~~~~~~~~~~ 
4 \frac{1}{(Y^2+\rho^2)^2} Y^{\mu}
Y^{\nu} h_{\mu \nu}(y) Y_a Y_b\bigr{]} 
\;\;.
\label{g'ab}
\end{eqnarray}
Note that in the $\rho \rightarrow 0$ limit, the leading term in
the metric deformation is in the Feffermann-Graham form, namely,
\be
\delta_{h} G'_{\rho \rho} \rightarrow 0\;\;,\;\;\;\;\;\;\delta_{h} G'_{\rho a} 
\rightarrow 0\;\;,\;\;\;\;\;\;
\delta_{h} G'_{ab} \rightarrow  \frac{1}{\rho^2} h_{ab}\;\;,
\label{fgl}
\ee
thus also in agreement with the predictions of the Groisser-Murray 
theorem discussed in section 2.3.

Having obtained $\delta_{h}G'_{AB}$, we could now attempt to verify the
linearized Einstein equation (\ref{leeq}) (with $\Lambda = -4$) by a
direct calculation.  However, we can side-step this calculation by noting
that the expressions (\ref{g'rhorho}) and (\ref{g'ab}) coincide exactly
with the $AdS_{5}$ boundary-to-bulk graviton propagator in the transverse
traceless gauge obtained in \cite{graviton}!  This proves that to ${\cal
O}(h)$, i.e.\ to first order in the metric perturbation, the deformation
of the Information Metric satisfies the bulk Einstein equation.

It is instructive (and necessary for the subsequent calculations)
to compute $\delta_{h} G_{AB}$ which appears naturally 
in the definition of the Information Metric.  For this we need to 
compute the diffeomorphism vector field $V_A$. These integrals can be 
performed using the above steps. The result is
\begin{eqnarray}
V_{\rho} &=& \frac{1}{\pi^2}\int d^4 y \frac{\rho}{(Y^2 +\rho^2)^4}
\bigl{[} 1 + \frac{2\rho^2}{Y^2+\rho^2}\bigr{]} Y^{\mu} Y^{\nu} h_{\mu
\nu}(y) \nonumber \\
V_a &=& \frac{2}{\pi^2} \int d^4 y \frac{\rho^2}{(Y^2 +\rho^2)^4}
\bigl{[} \delta_a^{\nu} - \frac{Y^{\nu} Y_a}{Y^2 + \rho^2}
\bigr{]} Y^{\mu}h_{\mu \nu}(y)\;\;.
\end{eqnarray}
Substituting $V_A$ and $\delta_{h} G'_{AB}$ in (\ref{dgab})
we finally obtain the following expressions for $\delta_{h} G_{AB}$:
\begin{eqnarray}
\delta_{h} G_{\rho \rho}&=&\frac{4}{\pi^2} \int d^4 y\frac{1}
{(Y^2+\rho^2)^4} Y^{\mu} Y^{\nu} h_{\mu \nu}(y)
\nonumber \\
\delta_{h} G_{\rho a} &=&\frac{6}{\pi^2}\int d^4 y \frac{\rho}{(Y^2+\rho^2)^4}
Y^{\mu} h_{\mu a}(y)
\nonumber \\
\delta_{h} G_{ab} &=& \frac{6}{\pi^2}\int d^4 y \frac{1}{(Y^2+\rho^2)^4}
\bigl{[} \rho^2  h_{ab}(y) -   
\frac{1}{3} \delta_{ab} Y^{\mu}
Y^{\nu} h_{\mu \nu}(y) \bigr{]} \;\;.
\label{grhoab}
\end{eqnarray}
$\d G_{AB}$ also (and somewhat more manifestly) satisfies (\ref{fgl}).

It is natural to ask what happens if the metric variation $h_{\mu\nu}$ is 
simply a diffeomorphism,
\be
h_{\mu\nu} = \del_{\mu}\xi_{\nu} + \del_{\nu}\xi_{\mu}\;\;,
\label{hmndx}
\ee
which acts non-trivially on the instanton density but which should not change the
bulk geometry. Indeed, by plugging (\ref{hmndx}) into (\ref{grhoab}), one finds
that 
\be
\d_{h}G_{AB} = \nabla_{A}X_{B} + \nabla_{B}X_{A}
\ee
is a bulk diffeomorphism, with $X_{\rho}=0$ and
\be
X_{a}(\rho,a) = \frac{6}{\pi^{2}}\int d^{4}y \frac{\rho^{2}}{(Y^{2}+\rho^{2})^{4}}
\xi_{a}(y)\;\;.
\ee
This is the natural lift of $\xi_{\mu}$ to $\M_{0}$ with 
\be
\lim_{\rho\ra 0} X_{a}(\rho,a) = \frac{1}{\rho^{2}}\xi_{a}(a)\;\;.
\ee

\subsection{The Hamilton-Jacobi Equation}

We now show that to the first order in the metric perturbation, the
Hamilton-Jacobi equation (\ref{b9}) which, as we have seen, says that
$\tr F^2$ is related to the (regularized) geodesic boundary-to-bulk
geodesic distance, still holds.  Taking the 
variation of (\ref{b9}), it follows  that we need to prove
\be
4 F^2(x) v_A(x) G^{AB} \delta_{h} v_B(x) = -2 F^2(x) v_A(x) v_B(x) 
\delta_{h} G^{AB}
\label{dhj}
\ee
where we have multiplied both sides by $2 F^2(x)$ for convenience.

This equation depends on $x$ and the moduli as well as the perturbation
$h_{\mu \nu}(y)$. Since the perturbation is arbitrary (with suitable
falloff conditions at large $y$), the above equation must be true for
each point $y$. Therefore we will drop the $y$-integrals in (\ref{df2})
and (\ref{grhoab}) in the following.
Note that $\delta_{h} F^2$, appearing on the left hand
side, has a singularity as $x$ approaches $y$ while the right hand side
has no such singularity. Thus the singularities on the left hand side
must cancel if the above equality is to hold. We will see below that
this is indeed the case.

First let us evaluate the right hand side. Using the expressions
(\ref{grhoab}) for $\delta_{h} G_{AB}$ one easily finds that
\begin{eqnarray}
-2 F^2(x) v_A(x) v_B(x) \delta_{h} G^{AB} &=&
\frac{48}{\pi^2 \rho^4}(F^2(x))^{\frac{3}{2}} F^2(y) 
 \bigl{[}\rho^4 X^{\mu}X^{\nu} +\rho^2 X^{\mu}Y^{\nu}
(X^2-\rho^2) + \nonumber \\ 
&~& ~~~~~~~
\frac{1}{6}Y^{\mu}Y^{\nu}(X^4-
4X^2\rho^2 +\rho^4) \bigr{]}h_{\mu \nu}(y)
\label{dhjr}
\end{eqnarray}

Now we proceed to compute the left hand side of (\ref{dhj}).  Using the
form of $\delta_{h} \tr F^2$ (\ref{df1}, \ref{df2}), the left hand side
can be written as
\begin{eqnarray}
4 F^2(x) v_A(x) G^{AB} \delta_{h} v_B(x) &=&\frac{1}{96} (v^A
\partial_A-4)
\delta_{h} \tr F^2 \nonumber \\
&=& \frac{1}{96}
(\partial_{\mu} \partial_{\nu} 
T^{\mu\nu} -2\partial_{\mu} T^{\mu} + T)
\label{dhjl}
\end{eqnarray}
where
\begin{eqnarray}
T^{\mu\nu} &=& (v^A(x) \partial_A -4) S^{\mu\nu} \nonumber \\
T^{\mu} &=& v^A_{\nu}(x) \partial_A S^{\mu \nu} \nonumber \\
T &=& v^A_{\mu\nu}(x) \partial_A S^{\mu \nu}
\label{Tmunu}
\end{eqnarray}
and $S^{\mu\nu}$ was defined in (\ref{df2}).
Here $v^A_{\mu}$ and $v^A_{\mu\nu}$ indicate derivatives of $v^A$ with 
respect to $x^{\mu}$,  $x^{\nu}$ etc. and the moduli space indices $A$, 
$B$ are raised and lowered with the $AdS_{5}$ metric $G_{AB}$. The 5-dimensional
tangent vector space at each point on $AdS_{5}$ is spanned by
the 5 vectors $v^A$ and $v^A_{\mu}$ (we will discuss this in more detail in
section 4.2). Therefore 
$v^A_{\mu \nu}$ can be expressed as a linear combination of $v^A$ and 
$v^A_{\lambda}$.  A short calculation shows that
\be
v^A_{\mu \nu}=-\delta_{\mu \nu}(4(F^2)^{1/2} v^A - z_{\lambda} v^A_{\lambda})
+(z_{\mu} v^A_{\nu} + z_{\nu} v^A_{\mu})
\label{vmn}
\ee
where $z_{\mu} = \frac{1}{4} \partial_{\mu} \ln F^2(x)$.
Using now the fact that $ S^{\mu}_{\mu} =0$, we obtain 
\be
T = 2z_{\mu} T^{\mu} 
\label{TTmu}
\ee
$T^{\mu \nu}$ and $T^{\mu}$ can be calculated explicitly using the 
expression (\ref{df2}) for  $S^{\mu \nu}$. 
It is convenient to write $S_{\mu \nu}$  as
\be
S_{\mu\nu}= -\frac{4}{\pi^2(x-y)^2}(F^2(x) F^2(y))^{\frac{3}{4}}
\hat{S}_{\mu\nu}
\ee
with
\be
\hat{S}^{\mu\nu} = \frac{1}{\rho^2} \tr (\rho^2 +Y\bar{X})\sigma^{\mu\lambda}
(\rho^2+X\bar{Y}) H_{\lambda}^{\;\nu}(y)
\ee
Then one can show that $\hat{S}_{\mu\nu}$ satisfies the following
identities:
\begin{eqnarray}
(v^A(x)\partial_A +2)\hat{S}_{\mu\nu}&=& (v^A(y)\partial_A
+2)\hat{S}_{\mu\nu}=0
\nonumber \\
(\Box - 12) \hat{S}_{\mu\nu}&=&0 
\label{dashat}
\end{eqnarray}

Using these identities it is straightforward to show that:
\begin{eqnarray}
T^{\mu \nu} &=& -\frac{4}{\pi^2 (x-y)^2}(F^2(x) F^2(y))^{\frac{3}{4}}
(v^A(x) \partial_A
+3v^A (x)(v_A(x) + v_A(y)) -4)\hat{S}^{\mu\nu}\nonumber\\
&=&6 \frac{\rho^2 (x-y)^2}{(\rho^2 +
X^2)(\rho^2+Y^2)} S^{\mu \nu}
\end{eqnarray}
where we have used (\ref{dashat}) and the explicit expression of
$v^A(x)v_A(y)$ given in (\ref{b16}).
The computation of $T^{\mu}$ is similar but
considerably more involved. The result is:
\be 
T^{\mu} = \frac{1}{3} (\partial_{\nu}-  z_{\nu}) T^{\mu \nu} 
\label{TmuTmunu}
\ee
Combining the equations (\ref{dhjl}), (\ref{Tmunu}), (\ref{TTmu}) and
(\ref{TmuTmunu}) we have
\begin{eqnarray}
4 F^2 v_A(x) G^{AB} \delta_{h} v_B(x)&=&\frac{1}{288}(\partial_{\mu} + 4z_{\mu})
\partial_{\nu} T^{\mu \nu}
\nonumber \\
&=& -\frac{1}{12\pi^2} F^2(x) F^2(y)  
(\partial_{\mu} + 8 z_{\mu})(\partial_{\nu}
+4 z_{\nu}) {\hat{S}}^{\mu\nu}
\label{dhjl1}
\end{eqnarray}
Using the identities (\ref{sigmai0}, \ref{sigmai1}) involving $\sigma$-matrices 
and their contractions with 
the anti-self-dual tensor $H$ given in (\ref{sigmai2}), one can show that
\begin{eqnarray}
\partial_{\mu} \partial_{\nu} \hat{S}^{\mu \nu} &=& -\frac{96}
{\rho^2} Y^{\mu} Y^{\nu}
h_{\mu \nu}(y)  
\nonumber\\
z_{\mu} \partial_{\nu} \hat{S}^{\mu \nu} &=& \frac{48}{\rho^2(\rho^2 + X^2)}
(X^2 Y^{\mu} Y^{\nu} + \rho^2 Y^{\mu} X^{\nu} ) h_{\mu \nu}(y) 
\nonumber\\
z_{\mu} z_{\nu} \hat{S}^{\mu \nu} &=& -\frac{16}{\rho^2(\rho^2 + X^2)^2} 
(X^2 Y^{\mu}+\rho^2 X^{\mu}) h_{\mu \nu}(y)
(X^2 Y^{\nu}+\rho^2 X^{\nu})
\label{dmushat}
\end{eqnarray}

Plugging these expressions into  (\ref{dhjl1}), one finds after a short
algebra that it reproduces the right hand side of (\ref{dhj}) given by (\ref{dhjr}).
We have thus established the validity of the HJ equation to ${\cal O}(h)$ in the
perturbation of the boundary metric.

We also want to point out that the calculation of this section suggests an alternative
strategy for determining the variation $\delta_{h}G_{AB}$ of the Information Metric. 
Namely, as the above calculation has shown, calculating $v^{A}\delta_{h}\va$ one
finds
\be
2v^{A}\delta_{h}\va = -\va\vc M^{AC}\;\;,
\label{altdg}
\ee
where $M^{AC}$ is an $x$-independent matrix. We can therefore define
a variation of $G_{AB}$ by, say, $\d G^{AB} = M^{AB}$. This metric
variation, by definition, preserves the HJ equation and will also preserve
the Propagator Equation. If one proceeds in this way (which is simpler
than the brute-force determination of $\d_{h}G_{AB}$ of section 3.2), then
one still has to show that indeed $M^{AB}=\d_{h}G^{AB}$, i.e.\
that $M^{AB}$ is what one would have obtained by varying the integral 
representation (\ref{i3}) of the Information Metric. 
We establish this in section 4.9.

\subsection{The Massless Boundary-to-Bulk Scalar Propagator}

We know that the boundary-to-bulk scalar propagator of AdS is simply
$\tr F^{2}$.
We will now show that to first order in $h$, $\tr F^2$ is still the 
boundary-to-bulk scalar propagator with respect to the bulk Information Metric. 
This amounts to proving
\begin{equation}
\Box \delta_{h} \tr F^2(x)  + (\delta_{h} \Box) \tr F^2(x) =0 
\label{dboxf2}
\ee
This equation depends on $x$ and the moduli as well as on the
perturbation $h_{\mu \nu}(y)$. As discussed above, this equation must
therefore be true at each point $x$, $y$ and at each point of the moduli space.
Note that $\delta_{h} \tr F^2$ in the first term on the left hand side  has a
$1/(x-y)^2$ pole while the second term has no such singularity. Thus for this
equation to hold the action of the AdS Laplacian should remove this 
singularity. 

Indeed, expanding $x$ around $y$ in the expression for $S_{\mu 
\nu}$ defined in (\ref{df2}), we find that the coefficient of the
leading singularity $1/(x-y)^2$ is proportional to $F^2(y)$ which
is annihilated by the AdS Laplacian. Similarly the first subleading
singularity of the form $(x-y)^{\mu}/(x-y)^2$ comes with a coefficient
which is given by a single $y$-derivative of $F^2(y)$ which is also
annihilated by the AdS Laplacian. 

Actually we need more than just the
cancellation of singularities for the above equation to work. In the
second term, the $x$-dependence only appears in $\tr F^2$ while the $y$-dependence 
appears only in $\delta_{h} G$ implicit in $\delta_{h} \Box$. This
means that the second term is a finite sum of factorized expressions
in $x$ and $y$. Quite remarkably, using the identities (\ref{dashat})
satisfied by $\hat{S}_{\mu\nu}$ one can show that
\begin{eqnarray}
\Box S_{\mu \nu}(x,y) &=& -\frac{4}{\pi^2}(F^2(x)F^2(y))^{\frac{3}{4}}(\Box
+ 6(v^A(x)+v^A(y))\partial_A -6 + 18 v^A(x)v_A(y))\hat{S}_{\mu\nu}
\nonumber \\
&=&36\frac{\rho^2 (x-y)^2}{(\rho^2 +
X^2)(\rho^2+Y^2)} S_{\mu \nu}(x,y) 
\end{eqnarray}
where in the first equality we have used (\ref{b9}) and (\ref{b11})
and in the second equality (\ref{dashat}) and (\ref{b16}).
Thus the first term also becomes a finite sum of factorized
expressions in $x$ and $y$. Using now the fact that
\be
(\delta_{h} \Box) \tr F^2(x) = \frac{1}{\sqrt{G}} \partial_A \delta_{h}
(G^{AB} \sqrt{G})\partial_B \tr F^2
\ee
and the metric variation (\ref{grhoab}), one can verify after some
algebra that (\ref{dboxf2}) is indeed true.

\section{The Information Geometry of the Perturbed Instanton Moduli Space $\M$}

\subsection{Preliminary Remarks}

In the previous sections we established three key results about the
Information Metric $G_{AB}$ on the metric-perturbed instanton moduli space, 
namely that
\begin{enumerate}
\item the Einstein Equation
\be
R_{AB}=-4G_{AB}
\label{E}\;\;,
\ee
\item
the Propagator Equation
\be
\Box F^{2} = 0\;\;,
\label{B}
\ee
\item and the Hamilton-Jacobi Equation 
\be
G^{AB}v_{A}(x)v_{B}(x) = 1
\label{A}
\ee
\end{enumerate}
hold to ${\cal O}(h)$ in the perturbation of the metric on $\RR^{4}$.

These results also imply, as for the flat metric, that to ${\cal O}(h)$
the massive boundary-to-bulk scalar propagator is $(F^{2})^{\Delta/4}$,
with
\be
\Box (F^{2})^{\Delta/4} = \Delta(\Delta-4)(F^{2})^{\Delta/4}\;\;.
\ee

While all this is very encouraging, evidently a number of things remain to be
understood. For example, so far we have no understanding to which extent
these three properties are independent of or dependent on each other. We 
also do not yet know if we can expect these equations to remain valid 
at second or higher order in the metric perturbation. 

To address these questions we will now explore the geometry of the
instanton moduli space implied by the equations (\ref{E}, \ref{B},
\ref{A}). We will see that it is in particular the (somewhat mysterious)
HJ equation (\ref{A}) which endows the moduli space with a very rich
geometrical structure.

As a consequence we will be able to show that 
\begin{itemize}
\item at first order in the metric perturbation 
the three fundamental equations are not independent: the HJ equation
(\ref{A}) and the propagator equation (\ref{B}) imply the Einstein equation
(\ref{E}); and that 
\item these three equations cannot hold simultaneously to quadratic order in the
metric perturbation. 
\end{itemize}

Throughout we will see that the geometry of the Information Metric
is interwoven in a subtle and beautiful way with the geometry of the
boundary space-time, and we believe that understanding these structures
will eventually lead to a better geometrical understanding of the metric
variation of the instantons themselves.

\subsection{Elementary Consequences of the Hamilton-Jacobi Equation}

The HJ equation
\be
G^{AB}v_{A}(x)v_{B}(x) = 1
\label{a1}
\ee
is quite remarkable as the right hand side is $x$-independent even though
the $x$-dependent quantities $\va(x)$ are contracted with the $x$-independent
Information Metric $G^{AB}$.

Since for the flat metric on $\RR^{4}$, the vector fields $\va(x)$
span the tangent space at each point of the moduli space $\M_{0}$,
they will continue to do so on the perturbed moduli space $\M$.

By differentiating the above equation with respect to $x$, we obtain 
\be
G^{AB}\del_{\mu}v_{A}(x) v_{B}(x) = 0\;\;.
\label{a2}
\ee
Hence the four gradient vector fields 
\be
\vam\equiv \del_{\mu}v_{A} =\del_{\mu}\del_{A}\log F^{2}
\ee
on $\M$ (labelled by $\mu$)
are orthogonal to $\va$. Moreover, for the flat metric on $\RR^{4}$ they
are mutually orthogonal (super- or subscripts $0$ will from now on 
always refer to the unperturbed instantons),
\be
G_{0}^{AB}\vam^{0}\vbn^{0} = 4 \d_{\mu\nu} (F_{0}^{2})^{1/2}\;\;,
\label{k0}
\ee 
and thus they are also linearly independent on the perturbed moduli space.
We will denote the non-degenerate scalar product of the $\vam$ by $K_{\mu\nu}$,
\be
K_{\mu\nu} = G^{AB}\vam\vbn\;\;.
\label{a3}
\ee
Note that 
\be
K^{0}_{\mu\nu}= 4 \d_{\mu\nu}\frac{\rho^{2}}{[(x-a)^{2}+\rho^{2}]^{2}}
\ee
is, up to the moduli-dependent coordinate transformation
\be
x^{\prime\mu}=(x^{\mu}-a^{\mu})/\rho \;\;,
\ee
simply the round metric (of radius 1) on the four-sphere in stereographic coordinates. 

Starting from the HJ equation, we have therefore produced a 
four-parameter family of frames (five linearly independent vector
fields) on $\M$. Actually, as all the vector fields are gradient
vector fields, these correspond to a four-parameter family of coordinate
systems on $\M$ - we will come back to this aspect of the story later.  

In any case it follows that the Information Metric on the moduli space
can be written purely algebraically as
\be
G_{AB} = \va(x)\vb(x)+ K^{\mu\nu}(x)\vam(x)\vbn(x)
\label{a9}
\ee
for any $x$. Once again it is a remarkable fact that the left hand side
of this equation is completely independent of $x$. 

Obviously, by construction, the integral and algebraic form of $G_{AB}$
are equivalent if the metric entering the HJ equation (\ref{a1}) is
the integral version of the Information Metric.  On the other hand if,
as proposed at the end of section 3.3, we define the variation of the
Information Metric directly through the variation of the HJ equation,
i.e.\ by (\ref{altdg}), then we still need to show that the variation
defined in this way is equal to the variation one would have obtained
by varying the integral version of the Information Metric. In order to
accomplish this, we need to accumulate some more facts about the geometry
implied by the HJ and propagator equations.  We will come back to this
issue in section 4.9.

\subsection{Properties of $K_{\mu\nu}$}

By differentiating the HJ equation once more, we obtain
\be
K_{\mu\nu} = - G^{AB}\del_{\nu}\vam \vb \;\;.
\label{a4}
\ee
To extract information from this equation, we proceed as follows
(similar arguments will be used repeatedly in the following).
Since the $\va$ and $\vam$ provide a basis for (co-)vectors on
$\M$ (for any $x$), in particular the 10 covectors $\del_{\nu}\vam$
can be expanded in this basis. Thus we have
\be
\del_{\nu}\vam(x)= C_{\mu\nu}^{\la}(x)\val(x) + D_{\mu\nu}(x)\va(x)
\label{a5}
\ee
for some coefficients $ C_{\mu\nu}^{\la}(x)$ and $D_{\mu\nu}(x)$.
Since $\del_{\nu}\vam(x)=\del_{\mu}\van(x)$, these coefficients 
are symmetric.

Plugging this expansion into (\ref{a3}) and using the orthogonality
(\ref{a2}) of $\va$ and $\vam$, one finds 
\be
D_{\mu\nu}(x) = -K_{\mu\nu}(x)\;\;.
\ee
Since $K_{\mu\nu}$ is a tensor with respect to $x$, it follows that 
$C_{\mu\nu}^{\la}$
is (transforms as) a connection, and we will denote the corresponding
covariant derivative by $D_{\mu}$. Hence we can write (\ref{a5}) as
\be
D_{\nu}\vam(x) = -K_{\mu\nu}(x)\va(x)\;\;.
\label{a6}
\ee
To see what $D_{\mu}$ is, we act with it on (\ref{a3}). Using (\ref{a2}),
one simply finds  
\be
D_{\la}K_{\mu\nu}=0\;\;.
\ee
Hence, since $K_{\mu\nu}$ is non-degenerate and $C_{\mu\nu}^{\la}$ is symmetric,
$D_{\mu}$ is the unique metric-compatible and torsion-free connection associated
with $K_{\mu\nu}$. For the flat metric, (\ref{a6}) reduces to (\ref{vmn}).

Using this, the innocuous-looking equation (\ref{a6}) has a somewhat surprising 
consequence. Namely, differentiating (\ref{a6}) once more and taking commutators,
one finds 
\be
[D_{\la},D_{\nu}]\vam = -(K_{\mu\nu}\val - K_{\mu\la}\van)\;\;.
\ee
Since the $\vam$, regarded as (a five-parameter family of) five vectors on $\RR^{4}$, 
span the tangent space at each $x$, it follows that the Riemann curvature tensor
of $K_{\mu\nu}$ has the form
\be
R(K)_{\sigma\nu\la\mu}=K_{\sigma\la}K_{\nu\mu}-K_{\sigma\mu}K_{\nu\la}
\;\;.
\ee
This characterizes the round metric on the four-sphere (of unit radius).
Therefore $K_{\mu\nu}$, for an arbitrary metric perturbation $h_{\mu\nu}$, can 
only differ from $K_{\mu\nu}^{0}$ by a coordinate transformation! 

This can also be checked explicitly. We need to show that the metric variation
of $K_{\mu\nu}$ is a diffeomorphism, i.e.\ that there exists a vector field 
$\xi_{\mu}$ such that
\be
\d_{h}K_{\mu\nu}= D_{\mu}\xi_{\nu} + D_{\nu}\xi_{\mu}\;\;.
\label{a7}
\ee
To analyse the metric variation of $K_{\mu\nu}$, as defined by (\ref{a3}), 
\be
\d_{h}K_{\mu\nu} = (\d_{h}G^{AB}) \vam\vbn + G^{AB}(\d_{h}\vam)\vbn
+ G^{AB}\vam\d_{h}\vbn\;\;,
\label{a8}
\ee
it is convenient to expand $\d_{h}\va$ in our standard basis as
\be
\d_{h}v_{A} = C\va + C^{\nu}\van\;\;.
\ee
Thus e.g.\
\be
C= G^{AB}\vb\d_{h}\va\;\;,
\ee
which, using $\d_{h}(G^{AB}\va\vb)=0$, can also be written as
\be
C=-\trac{1}{2}(\d_{h}G^{AB})\va\vb\;\;,
\ee
with similar expresions for $C^{\nu}$. 

Using
\be
\d_{h}\vam=\del_{\mu}\d_{h}\va
\ee
and (\ref{a6}), one finds that $\d_{h}\vam$ has the expansion
\be
\d_{h}\vam=(\del_{\mu}C - K_{\mu\nu}C^{\nu}) \va 
+ (D_{\mu}C^{\nu}+ \d_{\mu}^{\;\nu}C) \van\;\;.
\ee
Plugging this into (\ref{a8}) one finds after a short calculation 
that $\d_{h}K_{\mu\nu}$ indeed takes the form (\ref{a7}), with 
\be
\xi_{\mu}=-\trac{1}{2}\del_{\mu}C + K_{\mu\nu}C^{\nu} \;\;.
\ee

We thus obtain a map
\be
\mbox{Metric Deformations on $\RR^{4}$} \Ra 
\mbox{Coordinate Transformations on $S^{4}$} 
\ee
We will not pursue this matter here, but we believe that a better understanding of
this map (and appreciation of the raison d'\^etre of its existence) should lead to
a more geometrical understanding of the behaviour of instantons under metric
variations. 

Since the metric $K_{\mu\nu}$ is equivalent to the round metric on the four-sphere
for all values of the moduli $y^{A}$, it must also be true that 
\be
\del_{A}K_{\mu\nu} = D_{\mu}\eta_{A\nu} + D_{\nu}\eta_{A\mu}
\ee
for some vector field $\eta_{A\mu}$. This equation, which we will establish below,
once again displays the intricate way in which the geometry of the moduli space
interacts with that of the space-time.

To determine $\del_{A}K_{\mu\nu}$, 
we need to consider the covariant derivative $\nabla_{A}\vb$. As this is a 
(symmetric)
two-tensor on $\M$, it can be expanded as a bilinear in our basis $(\va,\vam)$. But
since
\be
v^{A}\nabla_{A}\vb = v^{A}\nabla_{B} \va =0\;\;,
\label{a10}
\ee
$\nabla_{A}\vb$ is orthogonal to $\va$ and $\vb$ and must therefore have an
expansion of the form 
\be
\nabla_{A}\vb = L^{\mu\nu}\vam\vbn\;\;,
\ee
where $L^{\mu\nu}$ is symmetric. For the flat metric one has
\be
L^{\mu\nu}_{0} = - K^{\mu\nu}_{0}\;\;,
\label{a11}
\ee
so that, using (\ref{a9}), 
we obtain the equation (\ref{b10}), namely
\be
\nabla_{A}^0\vb^0 = - G_{AB}^0 + \va^0\vb^0\;\;.
\label{a13}
\ee
Using the definition of $L_{\mu\nu}$, one can now determine 
\be
\del_{A}K_{\mu\nu}=-2L_{\mu\nu}\va + 
+\va^{\la} (D_{\mu}L_{\la\nu} + D_{\nu} L_{\la\mu})
\label{a14}
\ee
(where we have raised and lowered indices using the metric $K_{\mu\nu}$).
This essentially identifies $L_{\mu\nu}$ and its covariant derivatives 
(with respect to $x$) with components of the Christoffel symbols of the 
moduli space metric. We will make this correspondence more precise below.

Using (\ref{a6}), it now follows that $\del_{A}K_{\mu\nu}$ can be written
as
\be
\del_{A}K_{\mu\nu}=D_{\mu}(\va^{\la}L_{\la\nu}) +D_{\nu}(\va^{\la}L_{\la\mu})
\ee
so that
\be
\eta_{A\mu} = \va^{\la}L_{\la\mu}\;\;.
\ee

\subsection{The Curvature and the Role of the Propagator and Einstein Equations}
 
So far, everything we have derived is a consequence of the HJ equation
(\ref{A}) alone. We now come to the heart of the matter, namely the interplay
of the resulting geometry with the propagator and Einstein 
equations (\ref{B}, \ref{E}).

We first note that, given the HJ equation, the propagator equation is
equivalent to
\be
\Box F^{2} = 0 \LRa \nabla^{A}\va = -4 \LRa K_{\mu\nu}L^{\mu\nu}=-4\;\;.
\label{a16}
\ee
Note that for the flat metric this is identically satisfied because 
$L^{\mu\nu}_{0} = - K^{\mu\nu}_{0}$ (\ref{a11}).
 
Also, given the HJ equation the Einstein equation is equivalent to
\be
R_{AB}=-4G_{AB} \LRa R_{AB}v^{A}(x)v^{B}(x)=-4 \;\;\;\;\;\;\forall x\;\;.
\ee

Let us therefore calculate the Ricci tensor
\bea
R_{AB}v^{A}v^{B} &=& v^{B}[\nabla_{A},\nabla_{B}]v^{A} \non
&=& v^{B}\nabla_{A}\nabla_{B}v^{A} -v^{B}\nabla_{B}\nabla_{A}v^{A}\non
&=& -(\nabla_{A}v_{B})(\nabla^{A}v^{B}) -v^{B}\nabla_{B}\nabla_{A}v^{A} \;\;.
\eea
In passing to the last line we have used the symmetry $\nabla_{A}\vb=\nabla_{B}\va$ 
and (\ref{a10}). Expressing this in terms of $L^{\mu\nu}$, we obtain the key equation
\be
R_{AB}v^{A}v^{B} = - L^{\mu\nu}L_{\mu\nu} - v^{B}\nabla_{B} (L_{\mu\nu}K^{\mu\nu})
\label{a12}
\ee
relating the propagator and Einstein equations. 

We see that if the second term on the right hand side is zero, e.g.\ if
$\Box F^{2}=0$, then the Ricci tensor is negative (as a quadratic form).
To first order in the metric perturbation we can significantly sharpen
this statement to the statement that the metric is actually Einstein.

Indeed, since to zero'th order in the metric perturbation, i.e.\ for the flat
metric, we have (\ref{a11}), to ${\cal O}(h)$ in the metric
perturbation we can expand $L_{\mu\nu}$ as
\be
L_{\mu\nu} = -  K_{\mu\nu} + \ell_{\mu\nu} 
\ee
where $\ell_{\mu\nu}$ is of ${\cal O}(h)$ (we could alternatively have 
expanded $L_{\mu\nu}= -  K^{0}_{\mu\nu} + \tilde{\ell}_{\mu\nu}$
with $\tilde{\ell}_{\mu\nu}$ also of ${\cal O}(h)$).

Since we know that the propagator equation is true to ${\cal O}(h)$, 
we learn that to this order the second term in (\ref{a12}) is absent. We also
learn that $\ell_{\mu\nu}$ is traceless (with respect to either $K_{\mu\nu}$
or $K^{0}_{\mu\nu}$ - to this order this is the same),
\be
{\cal O}(h):\;\; K^{\mu\nu}L_{\mu\nu}=-4 \LRa K^{\mu\nu}\ell_{\mu\nu}=0\;\;.
\ee
But this now implies that
\be
L^{\mu\nu}L_{\mu\nu} = 4 + {\cal O}(h^{2})\;\;, 
\ee
so that
\be
R_{AB}v^{A}v^{B} = -4 + {\cal O}(h^{2})\;\;.
\ee
Thus we see that to ${\cal O}(h)$ the HJ and propagator equations imply
$L^{\mu\nu}L_{\mu\nu}=4$ and hence the Einstein equation!
Conversely, however, the HJ equation and the Einstein equation do not
imply the propagator equation.

Now let us assume, for argument's sake, that all three equations actually 
continue to hold to ${\cal O}(h^{2})$. In this case we can still expand 
$L_{\mu\nu}$ as before, and the propagator equation is satisfied if $\ell_{\mu\nu}$
is traceless with respect to $K_{\mu\nu}$. 
But now for the Einstein equation to hold to ${\cal O}(h^{2})$, we need 
\be
L^{2}=4 + {\cal O}(h^{3}) \;\;.
\ee
Thus the ${\cal O}(h)$-piece $\ell_{\mu\nu}^{1}$ of $\ell_{\mu\nu}$ has to be zero
(otherwise there would be a non-zero contribution to $L^{2}$ of ${\cal O}(h^{2})$),
and we obtain the result that 
\be
L_{\mu\nu} = -  K_{\mu\nu} + {\cal O}(h^{2})\;\;.
\ee
We will now show that this leads to a contradiction at ${\cal O}(h)$, namely that
this would imply that
the Information Metric is maximally symmetric for an arbitrary boundary metric
perturbation. 

Indeed, the same argument that leads to (\ref{a13}) shows that if 
$L_{\mu\nu}=-K_{\mu\nu}$ to some order, then
\be
\nabla_{A}\vb = - G_{AB} + \va\vb
\ee
to that order. But then the argument leading to (\ref{a18}) implies that
the bulk metric is maximally symmetric for an arbitrary ${\cal O}(h)$
variation of the boundary metric. Obviously for a number of reasons
this cannot be true - it is enough to think of symmetries and Killing
vectors. Alternatively, it follows from the Fefferman-Graham analysis
that a maximally symmetric bulk metric will induce a conformally flat
metric on the boundary - see \cite{ksss} for an explicit proof.

Thus we learn that not all three of our basic equations (\ref{E}, \ref{B}, \ref{A})
can be true to ${\cal O}(h^{2})$. 

\subsection{A Four-Parameter Family of Coordinate Systems on $\M$}

We have seen that it is quite useful to adopt the frame (actually, four-parameter
family of frames) $(\va,\vam)$ as a basis of (co-)tangent vectors on $\M$. This
can more succinctly be understood as a change of coordinates 
\be
y^{A}=(\rho,a^{\mu}) \ra z^{M}(y^{A})
\ee
on $\M$. Indeed, since $\va=\trac{1}{4}\del_{A}\log F^{2}$ and 
$\vam=\del_{A}\del_{\mu}\log
F^{2}$, the change of coordinates in question is
\be
(\rho,a^{\mu}) \ra (\trac{1}{4}\log F^{2}(x;\rho,a),
\trac{1}{4} \del_{\mu}\log F^{2}(x;\rho,a))\;\;.
\ee
Raising the index of $\del_{\mu}\log F^{2}$ with 
the boundary space-time metric $g_{\mu\nu}$, the four-parameter family of
coordinate transformation that we will actually consider is
\be
y^{A}=(\rho,a^{\mu}) \ra z^{M}(x) \equiv (r(x),z^{\mu}(x)) = 
(\trac{1}{4}\log F^{2}(x;\rho,a),\trac{1}{4}g^{\mu\nu}
 \del_{\nu}\log F^{2}(x;\rho,a))\;\;.
\ee

It follows from (\ref{a9}) that 
in these coordinates the Information Metric takes the simple form
\be
ds^{2}=G_{AB}dy^{A}dy^{B} = dr^{2}+\ga_{\mu\nu} dz^{\mu}dz^{\nu}\;\;,
\label{a17}
\ee
where
\be
\ga_{\mu\nu}=g_{\mu\rho}g_{\nu\la}K^{\rho\la}\;\;.
\label{a20}
\ee
The HJ equation $G_{AB}v^{A}v^{B}=1$ is now simply the statement
that $G_{r(x)r(x)}=1$ for all $x$ and (\ref{a2}) says that 
the off-diagonal component $G_{r(x)z^{\mu}(x)}=0$.

Note that if the propagator equation $\Box F^{2}=0$ holds in addition to the
HJ equation, then the functions $\exp 4r$ and $z^{\mu}$ are harmonic so
that these are harmonic coordinates for the Information Metric! This will be
useful later.

Note also that for the flat metric (and the choice $x=0$) 
the coordinate transformation
\be
(\rho,a^{\mu}) \ra (\ex{r_{0}}=(F_{0}^{2})^{1/4}, \trac{1}{8}
\del_{\mu}\log F_{0}^{2})
\ee
is just the inversion isometry
\be
(\rho,a^{\mu})\ra (\frac{\rho}{a^{2} + \rho^{2}}, \frac{a^{\mu}}{a^{2} + \rho^{2}})
\ee
of the standard AdS metric,
now more invariantly expressed in terms of $F_{0}^{2}$. This is reflected in the fact
that in the $z^{M}$-coordinates (for any value of $x$)
the unperturbed Information Metric takes the standard
AdS form
\be
ds_{0}^{2} = dr_{0}^{2} + \trac{1}{4}\ex{-2r_{0}}\d_{\mu\nu}dz^{\mu}dz^{\nu}\;\;.
\label{a15}
\ee

\subsection{The Geodesic Distance on $\M$ and the Instanton Density}

Writing the metric in the form (\ref{a15}) or (in general) (\ref{a17})
makes it manifest that $r = \frac{1}{4} \log F^{2}(x;\rho,a)$ is the
geodesic distance along paths of constant $z^{\mu}$.  It is also the
regularized geodesic distance between points with different values of
$z$ provided that for one of the end-points $r\ra\infty$ because the
difference between the $z$'s becomes irrelevant in this limit - in some
sense $r=\infty$ is a single point.

An alternative way of establishing this result is to use essentially 
{\em verbatim} the argument outlined in (\ref{gd1})-(\ref{gd2}) which
only relies on the validity of the equation $G^{AB}\va\vb=1$ which thus
once again can be interpreted as the HJ equation.

We have thus shown that to the order to which the HJ equation holds the
instanton density $F^{2}$ is directly related to the classical geodesic
distance of a boundary-to-bulk path in the geometry provided by the
Information Metric. One would dearly like to have a more conceptual
explanation for this.

\subsection{Gauss-Codazzi Equations for the Information Metric on $\M$}

Using the coordinates $(r,z^{\mu})$, equation (\ref{a14}) 
now has a much more transparent interpretation. In particular, since
\be
\del_{r} =v^{A}\del_{A}\;\;,
\ee
we have
\be
\del_{r}K_{\mu\nu} = -2L_{\mu\nu}\;\;.
\label{a19}
\ee
Thus $L_{\mu\nu}$ is essentially the extrinsic curvature (second fundamental form)
of $\ga_{\mu\nu}$, defined by
\be
\Theta_{\mu\nu} = \trac{1}{2}\del_{r}\ga_{\mu\nu}\;\;,
\ee
the precise relation being 
\be
\Theta_{\mu\nu} = g_{\mu\rho}g_{\nu\la}L^{\rho\la}\;\;.
\ee
For the unperturbed metric in the form (\ref{a15}) one evidently has
\be
\Theta_{\mu\nu}^{0} = - \ga_{\mu\nu}^{0}\;\;,
\ee
which is simply the equation $L_{\mu\nu}^{0}=-K_{\mu\nu}^{0}$ (\ref{a11}). 

The other component of (\ref{a14}), obtained by contracting $\del_{A}K_{\mu\nu}$
with $v^{A}_{\;\la}$ instead of $v^{A}$, then essentially says that the 
Christoffel symbols of $\ga_{\mu\nu}$, i.e.\ $z^{\la}$-derivatives of $\ga_{\mu\nu}$,
can be expressed as covariant $x$-derivatives of the extrinsic curvature 
$\Theta_{\mu\nu}$!
 
The Gauss-Codazzi equations express the curvature of the Information Metric 
$G_{MN}$ in terms of the curvature of $\ga_{\mu\nu}$ and the extrinsic curvature
$\Theta_{\mu\nu}$ and its trace $\Theta=\ga^{\mu\nu}\Theta_{\mu\nu}$. The expressions 
for the Ricci tensor are
\bea
R(G)_{rr} &=& -\del_{r}\Theta -\Theta^{\mu\nu}\Theta_{\mu\nu}\label{rr} \\
R(G)_{r\nu} &=& \nabla^{\mu}\Theta_{\mu\nu} - \del_{\nu}\Theta\label{rn}\\
R(G)_{\mu\nu}&=& R(\ga)_{\mu\nu}-\Theta\Theta_{\mu\nu} + 2
\Theta_{\mu\la}\Theta^{\la}_{\;\nu}-\del_{r}\Theta_{\mu\nu}\;\;.\label{mn}
\eea 
It follows that the scalar curvature of $G_{MN}$ is
\be
R(G) = R(\ga) - \Theta^{2} - \Theta_{\mu\nu}\Theta^{\mu\nu}-2\del_{r}\Theta\;\;.
\label{RG}
\ee

In particular, we now recognise the key equation (\ref{a12}) as nothing other 
than the $(rr)$-component (\ref{rr})
of the Gauss-Codazzi equation. Let us now analyse
these equations to first order in the metric perturbation. Since $\Theta_{\mu\nu}^{0}
=-\ga_{\mu\nu}^{0}$, we can expand $\Theta_{\mu\nu}$ like $L_{\mu\nu}$ as
\be
\Theta_{\mu\nu} =-\ga_{\mu\nu} + \theta_{\mu\nu}\;\;.
\ee
Given the HJ equation, the propagator equation $\Box F^{2}=0$ is equivalent to
$K_{\mu\nu}L^{\mu\nu}=-4$. In the present language this is the statement
\be
\Box F^{2}=0 \LRa \Theta = -4\;\;.
\ee
Thus we see that to first order in the metric perturbation, $\theta_{\mu\nu}$ 
(like $\ell_{\mu\nu}$) is traceless. Therefore to this order we also have
\be
\Theta^{\mu\nu}\Theta_{\mu\nu} = 4 + {\cal O}(h^{2})\;\;,
\ee
and we find that the $(rr)$-component of the Einstein equation 
\be
R_{rr} = -4G_{rr} = -4
\ee
is satisfied. Since it holds for all $x$ we know that all the other components
of the Einstein equation are also satisfied - this is just what we found before:
to ${\cal O}(h)$, the HJ and propagator equations imply the Einstein 
equation. 

However, we also learn one new fact about the curvature of $\ga_{\mu\nu}$.
Namely, using (\ref{RG}) and $R_{AB}=-4G_{AB}$, so that $R(G)=-20$, we see that  
\be
R(\ga)=0\;\;.
\ee
Thus to first order $\ga_{\mu\nu}$ is a scalar-flat perturbation of the flat 
(with respect to the $z^{\mu}$) metric $\ga_{\mu\nu}^{0}$. Note that,
even though it is true that (\ref{a17}) is Einstein if
\be
\ga_{\mu\nu}(r,z)=\ex{\pm 2r}\tilde{\ga}_{\mu\nu}(z)
\ee
with $\tilde{\ga}_{\mu\nu}(z)$ Ricci flat (hence the Fefferman-Graham expansion
is trivial in this case), there is no reason to expect the metric $\ga_{\mu\nu}$ 
to be Ricci-flat in general - there are simply not enough Ricci-flat perturbations
of the flat metric in Euclidean space. In fact there are none which are 
asymptotically flat while every transverse traceless perturbation of the flat metric
is automatically scalar flat - see (\ref{z5}) below.

Some further information about $\ga_{\mu\nu}$ and its curvature
$R(\ga)_{\mu\nu}$ can be obtained by looking at the other components of
the Gauss-Codazzi equations, but we will not pursue this here.

\subsection{The Relation between $\sqrt{K}$ and the Instanton Density}

In this section we will establish an identity which is needed to show 
directly the equivalence between the integral (\ref{i6}) and 
algebraic (\ref{a9}) forms of the Information Metric. The required identity is
\be
\sqrt{K} = 16 \sqrt{g}F^{2}\;\;.
\label{z1}
\ee
To zero'th order in the metric perturbation we indeed have the relationship
\be
\sqrt{K_{0}} = 16 F_{0}^{2}
\ee
(see (\ref{k0})).  We will now establish that this 
is yet one more equation which continues to hold to first order in the metric 
perturbation, i.e.\ that (\ref{z1}) is valid to  ${\cal O}(h)$.

First of all we note that the integrated version of (\ref{z1}),
\be
\int\sqrt{K} = 16 \int\sqrt{g}F^{2}\;\;,
\label{z2}
\ee
is automatically true. This follows because
$K_{\mu\nu}$ is equivalent to the round metric on the 
four-sphere for all metric perturbations  - see  (\ref{a7}). Thus we have
\be
\int\sqrt{K}=\int\sqrt{K_{0}}=\frac{8\pi^{2}}{3}\;\;.
\label{a21}
\ee
Moreover, and rather more obviously, 
\be
\int\sqrt{g}F^{2}=\int F_{0}^{2} = \frac{\pi^{2}}{6}\;\;,
\label{a22}
\ee
as the instanton number cannot change, i.e.\ $\d_{h} F^{2}$ is a total derivative. 

To establish (\ref{z1}), we proceed in three steps. We first show that
the $r$-dependence (i.e.\ $F^{2}$-dependence) of $\sqrt{K}$ is as given in
(\ref{z1}). This is easy.  We then show that $\sqrt{K}$ is independent of
the $z^{\mu}$. This requires a bit more work.  Lastly we show that the
normalization of $\int\sqrt{K}$ then fixes the possibly $x$-dependent
factor between the left- and right-hand sides of (\ref{z1}) to be equal
to the constant 16. This is once again easy.

To determine the $r$-dependence of $\sqrt{K}$, we use (\ref{a19}) which implies
\be
K^{\mu\nu}\del_{r}K_{\mu\nu}=-2K^{\mu\nu}L_{\mu\nu}\;\;.
\ee
Since the HJ equation holds to ${\cal O}(h)$, we have 
$K^{\mu\nu}L_{\mu\nu}=-4$ and thus
\be
\del_{r}\sqrt{K}=4\sqrt{K}\;\;,
\ee
which implies
\be
\sqrt{K}=\sqrt{g}F^{2}f(z,x)\;\;,
\ee
where $f(z,x)$ is some function of the coordinates $z^{\mu}$ and $x^{\mu}$.
We have introduced $\sqrt{g}$ on the right-hand-side  because $\sqrt{K}$ is
a density with respect to $x^{\mu}$ and any such scalar density can be written
as $\sqrt{g(x)}$ times a function. Alternatively, note that if we take our metric
perturbation to be traceless (by conformal invariance), then $\sqrt{g}=1$.

To determine the $z$-dependence, we make use of two facts: that the
$z^{\mu}$ are harmonic coordinates and that the scalar curvature
$R(\ga_{\mu\nu})=0$. We had already noted before that the propagator
equation implies that the $z^{\mu}$ are harmonic with respect to
the bulk $\Box$. Due to the special form (\ref{a17}) of the metric
in $(r,z)$-coordinates, this implies that the $z^{\mu}$ are also
harmonic with respect to the Laplacian $\Box^{\ga}$ associated to
$\ga_{\mu\nu}$. Indeed, denoting the Christoffel symbols of $G_{MN}$
and $\ga_{\mu\nu}$ by $\Gamma_{MNP}$ and $\ga_{\mu\nu\la}$ respectively,
we have
\be
\Box z^{\mu}=0 \LRa G^{NP}\Gamma^{\mu}_{\;NP}=0\LRa 
\ga^{\nu\la}\ga^{\mu}_{\;\nu\la}=0\LRa
\Box^{\ga}z^{\mu}=0\;\;.
\ee
Since the unperturbed metric $\ga_{\mu\nu}^{0}$ is the flat metric,
$\ga^{0}_{\mu\nu\la}=0$ and the harmonic gauge condition becomes
\be
\del_{z^\nu}\ga^{\;\nu}_{1\;\mu}=\trac{1}{2}\del_{z^\mu}\ga_{1}\;\;,
\ee
where indices have been raised and lowered with $\ga^{0}_{\mu\nu}$ and 
$\ga_{1}$ is the trace of the first-order perturbation $\ga^{1}_{\mu\nu}$
of $\ga^{0}_{\mu\nu}$.

In the harmonic gauge, the general expression for the linearized Ricci scalar
\be
R(\ga_{\mu\nu}) =
\del_{z^\mu}\del_{z^\nu}\ga_{1}^{\mu\nu}-
\ga_{0}^{\mu\nu}\del_{z^\mu}\del_{z^\nu}\ga_{1}
\label{z5}
\ee
becomes
\be
R(\ga_{\mu\nu}) = -\trac{1}{2}\ga_{0}^{\mu\nu}\del_{z^\mu}\del_{z^\nu}\ga_{1}\;\;.
\ee
Thus vanishing of the Ricci scalar implies 
\be
R(\ga_{\mu\nu})=0 \Ra \del_{z^\mu}\ga_{1} =0\;\;,
\ee
because for a small metric perturbation $\del_{z^\mu}\ga_{1}$ necessarily goes to zero
at infinity. 

To relate this to a statement about $K_{\mu\nu}$ we use (\ref{a20}) to deduce
\be
\del_{z^\mu}\ga_{1} = 0 \Ra \del_{z^\mu} K_{1}=0\;\;,
\ee
where $K_{1}$ is the trace of the first order perturbation (\ref{a7}),
\be
K^{1}_{\mu\nu}= D_{\mu}^{0}\xi_{\nu} + D_{\nu}^{0}\xi_{\mu}\;\;,
\ee
of $K_{\mu\nu}$. Hence
\be
\del_{z^\mu}\sqrt{K}=0
\ee
and
\be
\sqrt{K}=\sqrt{g}F^{2} f(x)\;\;.
\ee
Now to zero'th order we have $f(x)=16$. We thus write 
\be
f(x) = 16 + \phi(x) 
\ee
where $\phi(x)$ is of ${\cal O}(h)$. It now follows from (\ref{z2}) that
\be
\int F_{0}^{2}\phi(x) =0\;\;.
\ee
But since $F_{0}^{2}$ is the AdS boundary-to-bulk scalar propagator, this implies
that $\phi(x)=0$. Hence we have established (\ref{z1}). 

Incidentally note that this calculation can also be read as establishing
a relationship between the vector field $\xi^{\mu}$ appearing in the
variation of $K_{\mu\nu}$ under metric variations, and the vector field
$Y^{\mu}$ defined by
\be
\d_{h}F^{2} =\del_{\mu}Y^{\mu}\;\;.
\ee

\subsection{The Equivalence of the Integral and Algebraic Forms of the Information
Metric}

We have encountered two different forms of the Information Metric, the original
integral expression (\ref{i6}) which, in order to avoid confusion we will in this
subsection denote by $G_{AB}^{\mathrm int}$,
\be
G_{AB}^{\mathrm int}
= \frac{15}{8\pi^{2}}\int \sqrt{g}F^{2}\del_{A}\log F^{2}\del_{B}\log F^{2}
=\frac{30}{\pi^{2}}\int \sqrt{g}F^{2}\va\vb\;\;,
\ee
and the algebraic (vielbein) representation of the metric (\ref{a9}), i.e. 
\be
G_{AB} = \va\vb + K^{\mu\nu}\vam\vbn\;\;,
\ee
which could e.g.\ have been obtained by using (\ref{altdg}).
We now want to show directly that these two are equivalent. As $G_{AB}$ is
$x$-independent, we can write
\bea
G_{AB} &=& \frac{6}{\pi^{2}}\int\sqrt{g}F^{2}G_{AB}\non
&=& \frac{6}{\pi^{2}}\int\sqrt{g}F^{2}(\va\vb + K^{\mu\nu}\vam\vbn)\non
&=& \frac{1}{5}G_{AB}^{int} + \frac{6}{\pi^{2}}\int\sqrt{g}F^{2} K^{\mu\nu}\vam\vbn
\;\;.
\eea
We thus need to show that the second term equals $(4/5)G^{\mathrm int}_{AB}$. 
To proceed, we use (\ref{z1}) to rewrite this as
\be
\int\sqrt{g}F^{2} K^{\mu\nu}\vam\vbn
=\int\sqrt{K}K^{\mu\nu}\vam\vbn
\ee
We now integrate by parts to obtain
\be
\int\sqrt{g}F^{2} K^{\mu\nu}\vam\vbn = -\int\sqrt{g}F^{2} K^{\mu\nu}\va D_{\mu}\vbn
\;\;.
\ee
Using (\ref{a6}), we learn that 
\be
-\int\sqrt{g}F^{2} K^{\mu\nu}\va D_{\mu}\vbn = 4 \int\sqrt{g}F^{2}\va\vb\;\;.
\ee
Therefore this term gives precisely the missing contribution to $G^{\mathrm int}_{AB}$
and we have shown that
\be
G_{AB} = G_{AB}^{\mathrm int} \;\;.
\ee
To reiterate: this means that we do not have to define the variation
of the Information Metric through the variation of the $x$-integral
(\ref{i6}).  Instead we can extract the metric variation from the
variation (\ref{altdg}) of the HJ equation, use that expression to verify
the propagator (and thus the Einstein) equation, and we can prove (as
we just did) that the metric variation obtained in this (much simpler)
way is indeed the variation of (\ref{i6}).

\section{Conclusions}

In this paper we have shown that, to first order in the perturbation
of the boundary space-time metric, the Information Metric on
the 5-dimensional moduli space of $k=1$ $SU(2)$ instantons is
Einstein. Furthermore, to this order the perturbed instanton action
density is the corresponding boundary-to-bulk massless scalar propagator
and, quite remarkably, the regularized boundary-to-bulk geodesic
distance is proportional to the logarithm of the perturbed instanton
action density.

These results show that it is rather compelling to think of the bulk
space-time of the $AdS_{5}$/CFT${}_{4}$ correspondence as {\em being}
the instanton moduli space. Indeed, at least to this order, physically
relevant quantities of the non-trivial perturbed bulk space-time like
massless and massive scalar propagators and the geodesic distance are
then directly related to the simplest and most natural function on the
bulk space-time, namely the instanton density itself.

It would be nice to know if there is a similarly fruitful or suggestive
reinterpretation of the AdS/CFT bulk space-time in other dimensions.
For example, as for the unperturbed instanton moduli space, symmetry
arguments imply that the Information Metric on the moduli space of degree
one rational maps from the two-sphere to itself is the $AdS_{3}$-metric
\cite{murray}. In \cite{murray} it is also shown that the Information
Metric on the $(4k-1)$-dimensional space of rational maps of degree $k$
is non-degenerate for $k>1$. One can also ask if perhaps $AdS_7$
emerges as a moduli space of a 6-dimensional theory of interacting
anti-self-dual tensor fields.

The results we have obtained, although to a certain (limited) extent
anticipated from physics considerations, are certainly quite surprising
from a mathematical point of view. One would like to gain a better
mathematical insight in order to obtain more elegant proofs of these
statements. Note that the (rather cumbersome) proofs given in this paper
rely on the explicit form of the scalar instanton Green's functions.
One would like to be able to prove these statements in a more abstract
way, ideally by just making use of the self-duality equations. It would
also be good to have a deeper understanding of the significance of the
map from metric deformations on $\RR^{4}$ to diffeomorphisms of $S^{4}$
we obtained in section 4.3. Any such conceptual progress should help in
analyzing the higher order corrections and to see which of the above
statements, if any, continue to hold to higher orders in the metric
perturbation.

Since the only information we have at the moment is that to first order
the propagator and Hamilton-Jacobi equation imply that the Information
Metric is Einstein and that to second order not all the three statements
can be true simultaneously, it would also be useful to be able to analyze
the Information metric for a tractable (i.e.\ sufficiently symmetric)
metric on $S^{4}$ which is far from the (conformal class of the)
standard metric.

As we discussed in the introduction, the physics intuition which led us
to analyze this problem was based on the AdS/CFT correspondence together
with the D-instanton probe idea. However, our results are neither a test
nor a consequence of the AdS/CFT correspondence but rather logically
indepedent of it. In particular, there are two important differences
between this work and the usual AdS/CFT scenario. 

Firstly, the AdS/CFT correspondence is supposed to hold for ${\cal N}=4$
$SU(N)$ gauge theories in the large $N$ limit whereas here we are dealing
just with the $SU(2)$ theory.\footnote{Actually we were only considering
pure $SU(2)$ Yang-Mills because we were interested in the deformation of
the $AdS_5$ part of the bulk which only sees the bosonic moduli of the
super-instanton. However throughout this paper we have been assuming
an underlying ${\cal N}=4$ structure justifying the semi-classical
approximation around instanton solutions.} Secondly, the Information
Metric is certainly not the metric on the gauge fields that one normally
uses in studying Yang-Mills theories. The usual metric is the $L^2$-metric
which, as pointed out earlier, is not conformally invariant.

Perhaps the resolution to both the seeming differences lies in
taking the large $N$ limit where one integrates out all the zero modes
of the instanton associated to the gauge orientations (whose number
grows linearly with $N$), leaving behind just the $SU(2)$ instanton
moduli which in turn are identified with the bulk space-time \cite{dorey}. 

One way to test this would be to start with a 5-dimensional $SU(N)$ theory
reduced to one dimension on an instanton solution, with the moduli being
slowly varying functions of this coordinate. This reduction of course
gives the $L^2$-metric for the kinetic terms of the moduli. Integrating
out the moduli associated with gauge orientation should now give rise to
an effective kinetic term for the remaining moduli (namely those of the
$SU(2)$ instantons). The task would then be to check if in the large
$N$ limit this reduces to the Information Metric once all the other
fields have been integrated out, as might be expected on the basis of
4-dimensional conformal invariance. 

%Although suggestive, we do not see
%how the fact, explained at the end of section 2.2, that the Information
%Metric on the $SU(N)$ $k=1$ moduli space collapses to that for $SU(2)$
%could enter in this calculation.

If this is true, then one might start from an ${\cal N}=(4,4)$ sigma model
on the large $N$ $SU(N)$ instanton moduli space obtained by reducing
a 6-dimensional theory on a single instanton.  The large $N$ limit of
this sigma model should then give a definition of a string moving in
$AdS_{5}\times S^5$. This is somewhat reminiscent of the Matrix theory
description of the $(2,0)$-theory \cite{abkss} and related suggestions
in the AdS/CFT context \cite{mbhv}.

In any case, the results of this paper seem to suggest that the
one-instanton sector of $SU(2)$ gauge theory on a space which is
topologically $S^4$ or $\RR^{4}$ gives rise to a theory of gravity
on the instanton moduli space similar to the way that string theory
produces gravity on the target space. In particular, in string theory
the target space is also precisely the moduli (zero-mode) space of the
2-dimensional world-sheet theory.  Moreover, the criterion of conformal
invariance, which leads to the target space equations of motion in string
theory via the $\beta$-function equations, here led us to choose the
Information Metric on the moduli space which then turned out to satisfy
the (linearized) Einstein equations. In this sense, our construction
has some similarities with attempts to use Holographic Renormalization
Group ideas to (re-)construct the bulk space-time from the boundary
field-theory data - see e.g.\ \cite{hss,bgm}.

The Information Metric (or more precisely its first order deformation
$\d_{h}G_{AB}$) and $\int d^4 x \phi(x) \tr F^2(x)$ are certain operators
in the gauge theory whose semiclassical expectation values define the
on-shell linearized graviton and massless scalar field on the instanton
moduli space. Thinking along these lines, there are a number of questions
one might try to answer.

There are certainly infinitely many operators that one can construct
on the gauge theory side that are gauge and conformally invariant.
What other (perhaps massive) fields do they define in the bulk theory?
On the gauge theory side we can compute correlation functions of the
above operators beyond the semiclassical approximation. What would this
correspond to in the bulk theory? These quantities would be non-local on
the boundary but local in the instanton moduli space. Does this therefore
produce interaction vertices in the bulk theory? What is the role of
multi-instantons - multi-particle states?  What computation in the
gauge theory would yield information on quantum gravity effects in the
bulk? (We most certainly do not expect these to arise from a summation
over world-volume topologies as in string theory!) Etc. We believe that 
it is worthwhile to attempt to understand these issues.

\subsection*{Acknowledgements}

We would like to thank Edi Gava and Martin O'Loughlin for many useful
discussions.  This research is partially supported by EC contract no.\
CT-2000-00148.

\rnc{\Large}{\normalsize}

\end{document}